\documentclass[11pt,a4paper]{article}%
\usepackage{amsfonts}
\usepackage{amsmath}
\usepackage{amssymb}
\usepackage{graphicx}%
\providecommand{\U}[1]{\protect\rule{.1in}{.1in}}
\newtheorem{theorem}{Theorem}

\newtheorem{example}[theorem]{Example}

\def\RE{\mathbb{R}}

\begin{document}

\title{Boundaries and profiles in the Wigner formalism}
\author{Nuno Costa Dias\textbf{\thanks{Corresponding author; ncdias@meo.pt} }
\and Jo\~{a}o Nuno Prata\textbf{\thanks{joao.prata@mail.telepac.pt }}}
\maketitle

\begin{abstract}
We consider a quantum device contained in an interval in the context of the Weyl-Wigner formalism. This approach was originally suggested by Frensley, and is known to be plagued with several problems, such as non-physical and non-unique solutions. We show that some of these problems may be avoided if one writes the correct dynamical equation. This requires the (non-local) influence of the potential outside of the device and the inclusion of singular boundary potentials. We also discuss the problem of imposing boundary conditions on the Wigner function that mimic the effect of the external environment. We argue that these conditions have to be chosen with extreme care, as they may otherwise lead to non-physical solutions.
\end{abstract}


\subsection*{Declarations}

{\bf Funding:} Not applicable\\
{\bf Conflicts of interest/Competing interests:} Not applicable\\
{\bf Availability of data and material:} Not applicable\\
{\bf Code availability:} Not applicable\\

\section{Introduction}

The Weyl-Wigner formulation of quantum mechanics is an alternative formulation to the standard
formulation of Schr\"odinger and Heisenberg. Instead of wavefunctions or density matrices and self-
adjoint operators acting on Hilbert spaces, the Weyl-Wigner formulation is expressed in terms of quasi-distributions defined on the phase space and real c-number functions. Its formalism is close to classical 
statistical physics. Every state (pure or mixed) is associated to a Wigner function defined on the phase-space. Wigner functions were introduced by Wigner in \cite{Wigner}; they have many properties akin to a 
classical probability density for position and momentum. However, they are not necessarily everywhere 
non-negative. For this reason, they are often called a quasi-probability distribution. It was later realised that the expectation values of observables can be evaluated using the Wigner function and a c-number function on the phase space called the symbol of the operator. The correspondence between the 
operator and its symbol has been identified as Weyl's quantization rule \cite{Weyl}. 

The Weyl-Wigner formulation and Schr\"odinger's operator formulation can be shown to be completely 
equivalent \cite{Dias8}. Although this formulation is well developed and widely applied, there are 
surprisingly few articles discussing the question of boundaries in this context.

Boundaries are ubiquitous in physical systems. In particular, boundary value problems are one of the 
cornerstones of quantum mechanics. These are much more realistic models and moreover they entail a 
discretization of observables' spectra, which is one of the main features of quantum mechanics. They 
appear in virtually all branches of physics, ranging from quantum mechanics to general relativity, open 
strings and D-branes. In standard operator quantum mechanics some famous examples of boundary
value problems include the Kondo problem \cite{Affleck1}, quantum Hall liquids with constriction 
\cite{Moon}, and the Callan Rubakov model \cite{Affleck2}. Furthermore one of the simplest collision 
processes is the collision against an impenetrable boundary and some standard examples of quantum chaotic 
systems are bounded \cite{Lee,Lin}. 

In this work we wish to address certain boundary problems in the Wigner function formalism. This approach 
has become very popular for quantum device modeling and quantum transport 
\cite{Abdallah,Barletti,Degond,Dimov1,Ferry1,Frensley, 
Jacoboni,Jiang1,Jiang2,Jiang3,Manzini,Nedjalkov1,Rosati}.

In most articles, the authors choose to consider only the (integro-) differential equation governing the dynamics inside an interval, say, $\left[0,L\right]$, representing the quantum device, and discard the exterior of the device. The quantum device is hence regarded as a quantum open system. This strategy reduces the computational cost. However, the price to pay is that appropriate \textit{transparent} boundary conditions have to be imposed to account for the influence of the environment. In the present work we discuss two different methods to implement this strategy.  

In the first method, one starts by addressing the problem within the context of the Schr\"odinger equation. One then projects the solution onto the interval $\left[0,L\right]$ where the device is placed. The device becomes an open system with certain transparent boundary conditions. Subsequently, one applies the Weyl-Wigner transform to obtain the counterpart of this function in the phase space. The result is a true Wigner function supported on the same interval. However, this comes at a cost. Certain singular boundary potentials have to be imposed. To illustrate where this comes from, we discuss thoroughly the case of the particle in a box. This part of the present paper is based on previous results obtained by the authors in \cite{Dias1}-\cite{Dias6}. 

In the second method, one solves the Schr\"odinger equation on the whole line and then transforms it to the phase-space obtaining the associated Wigner function. After that, one projects it onto the interval of the device, and obtains a function which is supported on the infinite strip $\left[0,L\right] \times \mathbb{R}$. We will prove that the resulting function, which describes the device in the phase space, is never a Wigner function. But on the other hand, one is not forced to include the boundary potentials in the dynamical equations. 

In order to solve the dynamics directly in the phase space, using either of the two methods, one is led to impose certain boundary conditions at $x=0$ and $x=L$, which we will call \textit{profiles} of the Wigner function. We will discuss extensively, what kinds of functions are admissible as profiles of Wigner functions. 

Finally, as in the case of the second method, where one obtains solutions which are unphysical (in the sense that they are not Wigner functions), one may nevertheless look for the Wigner function which is closest to it. We discuss carefully how this can be done.

The problems related to the imposition of boundary profiles on Wigner functions have been studied in the past. For instance in \cite{Rosati} the authors consider the following "boundary profile problem" (BPP) for a phase space Wigner function $F(x,p)$:
\begin{equation}
[F,H]_M=0 \quad , \quad  \left\{
\begin{array}{l}
F(a,p)=g_a(p) \, , \, p>0 \\
F(b,p)=g_b(p) \, , \, p<0
\end{array}
\right.
\label{BPP}
\end{equation}
where $[\, , \, ]_M$ is the Moyal bracket, and $a<
b$ are two points where $F$ should satisfy a profile compatible with inflow boundary conditions, given by $g_a$ and $g_b$, respectively. Moreover $H$ is a Hamiltonian (of a specific type). The authors investigate whether: 1) there are solutions (Wigner functions or not) of this BPP; 2) there are solutions which are Wigner functions and 3) these solutions are unique. They conclude that a solution always exists but it might not be a Wigner function and may not be unique. The paper also discusses what happens in the case of a dissipative/decoherence dynamics obtained by adding a non-linear term to the Moyal equation. 

The same type of problems and possible approaches used to eliminate unphysical and non-unique solutions where discussed in other papers (e.g. \cite{Dimov1,Lotti}).

It is important to distinguish these problems from the ones we will consider here. In sections 3,4 and 5 we will be concerned not with boundary conditions (or boundary profiles) but mainly with {\it localization}. We want to study if a localized phase space function (understood as a function with support on an infinite strip $S=[a,b] \times \RE$) can be a Wigner function, and what type of $*$-genvalue and Moyal equations does it satisfy. We show that the localization of Wigner functions requires that distributional boundary potentials be added to the Hamiltonian. We also show that the projection of a globally defined Wigner function onto the strip $S$ is never a Wigner function, even after an obvious normalization (unless the support of the original Wigner function is already in $S$). These two results are largely independent of the boundary conditions to be satisfied by the Wigner functions (although these may affect the specific form of the boundary potentials - see \cite{Dias6}). 

The imposition of boundary profiles is a complementary problem to that of localization. In this paper we discuss the issues related to the boundary profiles in section 6. However, we will focus on a pure {\it kinematical} problem (and not on the BPP (\ref{BPP}) considered in \cite{Rosati}). We don't assume that the Wigner function is a solution of the Moyal equation or of the $*$-genvalue equation for some specific Hamiltonian; instead the aim is to identify the one-point profiles $g_a(p)$ (and the two-point profiles $g_a(p)$ and $g_b(p)$) for which there is at least one (pure or mixed) Wigner function $F(x,p)$ solution of the equations 
$$
F(a,p)=g_a(p) \quad \mbox{or} \quad \left\{ 
\begin{array}{l}
F(a,p)= g_a(p)\\
F(b,p)=g_b(p)
\end{array}
\right. ~.
$$
The focus is in obtaining explicit characterisations of the profiles which are compatible with Wigner functions (Wigner profiles). Quite interestingly there is a lot of freedom (in the choice of $g_a(p)$) if we impose just one profile on $F$; but if we impose two profiles then we find that they have to satisfy extremely restrictive conditions. If, in addition, the Wigner functions are solutions of some specific dynamical equations, then the conditions on the profiles will be even more restrictive. The results of this section are quite general and, to the best of our knowledge, they are new.

\section{Wigner functions}

Given a function $\psi \in L^2(\mathbb{R}^d)$ its Wigner distribution is defined as:
\begin{equation}
\begin{array}{c}
W \psi(x,p)= \frac{1}{(2\pi \hbar)^d}\int_{\mathbb{R}^d} \psi\left(x+\frac{y}{2} \right) \psi^{\ast} \left(x-\frac{y}{2} \right) e^{- \frac{i}{\hbar}p \cdot y} dy=\\
\\
=\frac{1}{(\pi \hbar)^d}\int_{\mathbb{R}^d} \psi\left(x+y \right) \psi^{\ast} \left(x-y \right) e^{- \frac{2i}{\hbar}p \cdot y} dy~.
\end{array}
\label{eqWignerFunctions1}
\end{equation}
Let us briefly review some of the most relevant properties of a Wigner function. 
It is a real function and, if $\|\psi\|_{L^2 (\mathbb{R}^d)}=1$, it is normalized:
\begin{equation}
\int_{\mathbb{R}^d} \int_{\mathbb{R}^d} W \psi(x,p)dx dp=1~.
\label{eqWignerFunctions2}
\end{equation}
It also has the proper marginal distributions:
\begin{equation}
\int_{\mathbb{R}^d} W \psi(x,p)dp = | \psi(x)|^2 \geq 0, \hspace{0.5 cm} \int_{\mathbb{R}^d} W \psi(x,p)dx = |\widetilde{\psi} (p)|^2 \geq 0~.
\label{eqWignerFunctions3}
\end{equation}
Here, $\widetilde{\psi}$ denotes the Fourier transform of $\psi$:
\begin{equation}
\widetilde{\psi}(p)= \frac{1}{(2\pi \hbar)^{d/2}} \int_{\mathbb{R}^d} \psi(x) e^{- \frac{i}{\hbar} x \cdot p } dx~.
\label{eqWignerFunctions4}
\end{equation}
Moyal's identity holds for all $\psi, \phi \in L^2 (\mathbb{R}^d)$:
\begin{equation}
\int_{\mathbb{R}^d}dx \int_{\mathbb{R}^d} dp~ W \psi(x,p) W \phi(x,p)= \frac{1}{(2 \pi \hbar)^d}\left| \langle \phi | \psi \rangle \right|^2~.
\label{eqWignerFunctions4Moyal}
\end{equation}

From the Cauchy-Schwarz inequality, one can show that Wigner functions are uniformly bounded
\begin{equation}
\left| W \psi(x,p) \right| \leq \frac{1}{(\pi \hbar)^d} ~, \text{ for all } x,p \in \mathbb{R}^d~,
\label{eqWignerFunctions4A}
\end{equation}
and uniformly continuous in $\mathbb{R}^{2d}$.

Very often, we need to consider the Wigner function associated with the density matrix of a mixed state:\begin{equation}
\left(\widehat{\rho} \psi\right)(x)= \int_{\mathbb{R}^d} \rho(x,y) \psi(y) dy~,
\label{eqWignerFunctions13}
\end{equation}  
with kernel
\begin{equation}
\rho(x,y)= \sum_{j} p_j \psi_j (x) \psi_j^{\ast} (y)~, 
\label{eqWignerFunctions14}
\end{equation}  
where $(p_j)$ is some probability distribution. The Wigner function is:
\begin{equation}
W \rho(x,p)=\frac{1}{(\pi \hbar)^d} \sum_{j} p_j  \int_{\mathbb{R}^d} \psi_j\left(x+y \right) \psi_j^{\ast} \left(x-y \right) e^{- \frac{2i}{\hbar}p \cdot y} dy~.
\label{eqWignerFunctions5A}
\end{equation}

In general\footnote{A notable exception are Gaussian states.}, it is very difficult to assess, whether a given phase space function $F$ is the Wigner function of some density matrix. It can be shown \cite{Dias8} that $F(x,p)$, defined on the phase space $\mathbb{R}^{2d}$, is the Wigner function associated to some density matrix if and only if:

\begin{enumerate}
\item $F$ is real,

\item $F$ is continuous and $F \in L^2 (\mathbb{R}^{2d})$,

\item $F$ is normalized, i.e. $F$ satisfies (\ref{eqWignerFunctions2}),

\item $F$ satisfies the positivity condition
\begin{equation}
\int_{\mathbb{R}^d} \int_{\mathbb{R}^d} F(x,p) W \psi(x,p) d x dp \geq 0~,
\label{eqWignerFunctions5B}
\end{equation}
for all $\psi \in L^2(\mathbb{R}^d)$.
\end{enumerate}

The latter condition is somewhat tautological, as it requires the knowledge of the entire set of Wigner functions associated to pure states. An alternative formulation suggested by Narcowich \cite{Narcowich} dispenses pure states.

A starting point in dealing with the positivity condition (\ref{eqWignerFunctions5B}) would be some condition for the function $F$ to be the Wigner function of a pure state. Such a condition was suggested by Tatarskii \cite{Tatarskii}. Let us define
\begin{equation}
\mathcal{Z}(x, \tau)=\int_{\mathbb{R}^d}F(x,p)e^{i p \cdot \tau} dp~.
\label{eqWignerFunctions5C}
\end{equation}
Then $F$ is the Wigner function associated with a pure state if and only if $F$ is real, normalized and the following \textit{wave equation} holds:
\begin{equation}
\frac{\partial^2}{\partial \tau^2} \ln \mathcal{Z}(x, \tau) - \left(\frac{\hbar}{2}\right)^2 \frac{\partial^2}{\partial x^2} \ln \mathcal{Z}(x, \tau)=0~.
\label{eqWignerFunctions5D}
\end{equation}

\subsection{The support property}\label{SubsectionSupportProperty}

Since we will be mainly interested in the question of boundaries, it is interesting to remark that if $\psi(x) = 0$, for $x_j < a_j$ (resp. $x_j >a_j$), then we also have:
\begin{equation}
W \psi(x,p)= 0, \text{ for all } p \in \mathbb{R}^d \text{ and } x_j < a_j \text{ (resp. }x_j >a_j~)
\label{eqWignerFunctions5}
\end{equation}  
In particular, if $\psi$ is supported in a hypercube of length $L$,
\begin{equation}
\psi(x)=0 , \text{for all }x \notin Q_L= \left\{x\in \mathbb{R}^d: ~ - \frac{L}{2} \leq x_j \leq \frac{L}{2}, \, \forall j=1,..d \right\}~,
\label{eqWignerFunctions6}
\end{equation}  
then 
\begin{equation}
W \psi (x,p)=0 , \text{ for all }x \notin Q_L= \left\{x\in \mathbb{R}^d: ~ - \frac{L}{2} \leq x_j \leq \frac{L}{2},\, \forall j=1,..d \right\}~ ,
\label{eqWignerFunctions7}
\end{equation} 
and all $p \in \RE$. The reciprocal of previous result also holds true. Thus, if $W\psi(x,p)$ vanishes for all $p \in \mathbb{R}^d$ when $x_j < a_j$, then $\psi(x)$ vanishes for $x_j < a_j$.

Let us briefly sketch the proof of this in dimension $d=1$. Suppose that $\psi(x)=0 $ for all $x <0$. Then $\psi(x+y) \psi^{\ast} (x-y)$ vanishes for all $x,y$ such that
\[
x-y <0 ~ \text{ or } ~ x+y <0 \Leftrightarrow y>x ~ \text{ or } ~ y <-x~.
\]
and hence $W \psi (x,p)$, vanishes for all $x<0$ and all $p \in \mathbb{R}$. 

Conversely, suppose that $W \psi(x,p)=0$ for all $x<0$ and all $p \in \mathbb{R}$. From the marginal property (\ref{eqWignerFunctions3}), we conclude that $|\psi(x)|^2 = \int_{\mathbb{R}} W \psi(x,p) dp=0$ for all $x<0$. Consequently $\psi(x)=0$, for all $x<0$.

\subsection{The Weyl transform}

The Wigner function (\ref{eqWignerFunctions1}) can be regarded as the Weyl transform of the rank-one operator $|\psi><\psi|$. Specifically, let $\widehat{A}$ denote some linear operator with (distributional) kernel $K_A(x,y)$ acting on the Hilbert space $L^2 (\mathbb{R}^d)$:
\begin{equation}
\left(\widehat A \psi\right)(x)= \int_{\mathbb{R}^d} K_A(x,y) \psi(y) dy~.
\label{eqWignerFunctions8}
\end{equation}  
The associated Weyl transform is given by:
\begin{equation}
\left(\mathcal{W}\widehat{A} \right)(x,p)= \int_{\mathbb{R}^d} K_A \left(x+\frac{y}{2},x-\frac{y}{2}\right) e^{-\frac{i}{\hbar} y \cdot p} dy~. 
\label{eqWignerFunctions9}
\end{equation}  
The Weyl transform of an operator is usally called the \textit{Weyl symbol} of the operator. The operator $|\psi><\psi|$ acts on $|\phi>$ as $< \psi|\phi>~|\psi>$. Hence, in the position representation, it has kernel $K(x,y)=\psi(x) \psi^{\ast}(y)$:
\begin{equation}
|\psi><\psi| \phi (x) = \int_{\mathbb{R}^d} \psi(x) \psi^{\ast}(y) \phi(y) dy= <\psi|\phi> \psi(x)~.
\label{eqWignerFunctions10}
\end{equation}  
We thus get that:
\begin{equation}
\begin{array}{c}
\frac{1}{(2\pi \hbar)^d} \left(\mathcal{W}|\psi><\psi| \right)(x,p)=\\
\\
= \frac{1}{(2\pi \hbar)^d}\int_{\mathbb{R}^d} \psi\left(x+\frac{y}{2}\right) \psi^{\ast} \left(x-\frac{y}{2}\right) e^{- \frac{i}{\hbar} y \cdot p} dy= W\psi(x,p)~.
\end{array}
\label{eqWignerFunctions11}
\end{equation}  
By analogy, we can construct the non-diagonal Wigner functions:
\begin{equation}
\begin{array}{c}
\frac{1}{(2 \pi \hbar)^d} \left(\mathcal{W}|\psi><\phi| \right)(x,p)=\\
\\
= \frac{1}{(2 \pi \hbar)^d}\int_{\mathbb{R}^d} \psi\left(x+\frac{y}{2}\right) \phi^{\ast} \left(x-\frac{y}{2}\right) e^{- \frac{i}{\hbar} y \cdot p} dy= W(\psi,\phi)(x,p)~.
\end{array}
\label{eqWignerFunctions12}
\end{equation}  
We may also consider density matrices (\ref{eqWignerFunctions13},\ref{eqWignerFunctions14}).

The corresponding Wigner function is:
\begin{equation}
\begin{array}{c}
W \rho(x,p)= \frac{1}{(2\pi \hbar)^d} \left(\mathcal{W}\widehat{\rho} \right)(x,p)\\
\\
= \frac{1}{(2\pi \hbar)^d}\int_{\mathbb{R}^d} \rho\left(x+\frac{y}{2},x-\frac{y}{2}\right) e^{- \frac{i}{\hbar} y \cdot p} dy \\
\\
=\frac{1}{(2\pi \hbar)^d}\sum_{j} p_j \int_{\mathbb{R}^d} \psi_j\left(x+\frac{y}{2}\right) \psi_j^{\ast} \left(x-\frac{y}{2}\right) e^{- \frac{i}{\hbar} y \cdot p} dy~.
\end{array}
\label{eqWignerFunctions15}
\end{equation}  
The Weyl transform can also be applied to unbounded operators. For instance, the kernel of the position operator $\widehat{X}_j$ is $ x_j \delta^d(x-y)$, were $ \delta^d(x-y)= \prod_{k=1}^d \delta(x_k-y_k)$. Indeed:
\begin{equation}
\left(\widehat{X}_j \psi \right) (x) = \int x_j \delta^d(x-y) \psi(y) dy= x_j \psi(x)~.
\label{eqWignerFunctions16}
\end{equation}  
Consequently, its Weyl transform is:
\begin{equation}
\begin{array}{c}
\left(\mathcal{W} \widehat{X}_j \right)(x,p)= \int \left(x_j+\frac{y_j}{2}\right) \delta^d \left[x+ \frac{y}{2}- \left(x- \frac{y}{2} \right) \right] e^{- \frac{i}{\hbar} p \cdot y }dy = \\
\\
= \int \left(x_j+\frac{y_j}{2}\right) \delta^d (y)e^{- \frac{i}{\hbar} p \cdot y }dy= x_j
\end{array}
\label{eqWignerFunctions17}
\end{equation}  
Likewise the momentum operator $\widehat{P}_j$ has kernel $ i \hbar \frac{\partial}{\partial y_j} \delta^d (x-y)$:
\begin{equation}
\begin{array}{c}
\left(\widehat{P}_j \psi \right)(x)= i \hbar \int \frac{\partial}{\partial y_j} \delta^d (x-y) \psi(y) dy =\\
\\
=-i \hbar \int \delta^d (x-y)  \frac{\partial}{\partial y_j} \psi(y) dy
 =-i \hbar  \frac{\partial}{\partial x_j} 	\psi (x)~.
\end{array}
\label{eqWignerFunctions18}
\end{equation}  
Let $\delta_j^{\prime d }(x)$ denote the partial distributional derivative with respect to the $j$-th coordinate. Then, the Weyl symbol of $\widehat{P}_j$ is:
\begin{equation}
\begin{array}{c}
\left(\mathcal{W} \widehat{P}_j \right)(x,p)= -i \hbar\int \delta_j^{\prime d}\left[x+ \frac{y}{2}- \left(x- \frac{y}{2} \right) \right] e^{- \frac{i}{\hbar} p \cdot y }dy=\\
\\
= -i \hbar\int \delta_j^{\prime d }(y) e^{- \frac{i}{\hbar} p \cdot y }dy=i \hbar\int \delta^{d }(y)\left(- \frac{i}{\hbar} p_j \right) e^{- \frac{i}{\hbar} p \cdot y }dy=p_j ~.
\end{array}
\label{eqWignerFunctions19}
\end{equation} 

\subsection{The star-product}

Given two operators $\widehat{A}$ and $\widehat{B}$, with Weyl symbols $A(x,p)= \mathcal{W}\widehat{A}$
and $B(x,p)= \mathcal{W}\widehat{B}$, respectively, the Weyl symbol of their product $\widehat{A} \widehat{B}$ can be expressed in terms of $A$ and $B$ by way of the so called \textit{star-product} (or Moyal product):
\begin{equation}
\mathcal{W}\left(\widehat{A} \cdot \widehat{B}\right) = A \star B~,
\label{eqWignerFunctions20}
\end{equation} 
where
\begin{equation}
\begin{array}{c}
\left(A \star B \right) (x,p)=\frac{1}{(\pi \hbar)^{2d}} \int \int \int \int A(x^{\prime},p^{\prime}) B (x^{\prime \prime},p^{\prime \prime}) \times \\
\\
\times e^{-\frac{2i}{\hbar}\left[p \cdot(x^{\prime}-x^{\prime \prime})+p^{\prime} \cdot (x^{\prime \prime}-x)+ p^{\prime \prime } \cdot (x-x^{\prime}) \right]}dx^{\prime} dx^{\prime \prime} dp^{\prime} dp^{\prime \prime}~.
\end{array}
\label{eqWignerFunctions21}
\end{equation} 
If either $A$ or $B$ is a polynomial, then the star-product can be written as an expansion in powers of Planck's constant:
\begin{equation}
\begin{array}{c}
\left(A \star B \right) (x,p)=A(x,p) B(x,p) + \frac{i \hbar}{2} A \overleftrightarrow{\mathcal{P}} B +\\
\\
+\frac{1}{2 !} \left(\frac{i \hbar}{2}\right)^2 A \overleftrightarrow{\mathcal{P}}^2 B+\frac{1}{3!} \left(\frac{i \hbar}{2}\right)^3 A \overleftrightarrow{\mathcal{P}}^3 B+ \cdots~,
\end{array}
\label{eqWignerFunctions22}
\end{equation} 
where $\overleftrightarrow{\mathcal{P}}$ is the Poisson operator:
\begin{equation}
 \left(A \overleftrightarrow{\mathcal{P}} B\right) (x,p)= \left\{A(x,p),B(x,p)\right\}= \sum_{j=1}^d \left(\frac{\partial A}{\partial x_j} \frac{\partial B}{\partial p_j}-\frac{\partial A}{\partial p_j} \frac{\partial B}{\partial x_j}\right)~,
\label{eqWignerFunctions23}
\end{equation} 
and $\left\{ \cdot, \cdot \right\}$ is the Poisson bracket.

In the same vein, we obtain the Weyl symbol of the quantum commutator, which is commonly known as the \textit{Moyal bracket}:
\begin{equation}
\begin{array}{c}
\left[A(x,p), B(x,p) \right]_M= \frac{1}{i \hbar} \mathcal{W} \left(\left[\widehat{A},\widehat{B}\right]\right)=\\
\\
= \frac{2}{\hbar (\pi \hbar)^{2d}} \int \int \int \int A(x^{\prime},p^{\prime})B(x^{\prime \prime}, p ^{\prime \prime}) \times \\
\\
\times \sin \left\{\frac{2}{\hbar} \left[p\cdot(x^{\prime}-x^{\prime \prime})+p^{\prime}\cdot (x^{\prime \prime}-x) + p^{\prime \prime } \cdot (x-x^{\prime})\right] \right\} dx^{\prime} d x^{\prime \prime } d p^{\prime} d p^{\prime \prime}~.
\end{array}
\label{eqWignerFunctions24}
\end{equation} 
Again, if either $A$ or $B$ is a polynomial, we obtain the expansion:
\begin{equation}
\left[A(x,p), B(x,p) \right]_M= \left\{A(x,p),B(x,p) \right\} -\frac{\hbar^2}{24} A \overleftrightarrow{\mathcal{P}}^3 B+ \cdots~.
\label{eqWignerFunctions25}
\end{equation} 
Thus, Weyl quantization amounts to an $\hbar$ deformation of the pointwise product (\ref{eqWignerFunctions22}) and of the Poisson bracket (\ref{eqWignerFunctions25}). This is the reason why it was, historically, the starting point for a quantization programme known as \textit{deformation quantization} \cite{Bayen}.

\subsection{The stargenvalue equation}

Let $|a>$ and $|b>$ (belonging to the Hilbert space) be some eigenstates of the Hermitian operator $\widehat{A}$ with eigenvalues $a$ and $b$, respectively:
\begin{equation}
\widehat{A}|a>=a|a>~, \hspace{1 cm} \widehat{A}|b>=b|b>~.
\label{eqWignerFunctions26}
\end{equation}
Consequently, we have:
\begin{equation}
\widehat{A}|a><b|=a|a><b|~, \hspace{1 cm} |a><b| \widehat{A}=b|a><b|~.
\label{eqWignerFunctions27}
\end{equation}
If $\psi_a(x)=<x|a>$ and $\psi_b(x)=<x|b>$, then the Weyl symbol of $|a><b|$ is (up to a normalization) the non-diagonal Wigner function (\ref{eqWignerFunctions12})  
\begin{equation}
\begin{array}{c}
\frac{1}{(2 \pi \hbar)^d} \left(\mathcal{W}|a><b| \right)(x,p)=\\
\\
= \frac{1}{(2 \pi \hbar)^d}\int_{\mathbb{R}^d} \psi_a\left(x+\frac{y}{2}\right) \psi_b^{\ast} \left(x-\frac{y}{2}\right) e^{- \frac{i}{\hbar} y \cdot p} dy= W(\psi_a,\psi_b)(x,p)~.
\end{array}
\label{eqWignerFunctions28}
\end{equation}  
If $A$ is the Weyl symbol of $\widehat{A}$, then we obtain from (\ref{eqWignerFunctions20},\ref{eqWignerFunctions27},\ref{eqWignerFunctions28}) the stargenvalue equations:
\begin{equation}
A \star W(\psi_a,\psi_b)=a W(\psi_a,\psi_b)~, \hspace{1 cm} W(\psi_a,\psi_b) \star A= b W(\psi_a,\psi_b)~.
\label{eqWignerFunctions29}
\end{equation}  
In particular for the diagonal Wigner function $W \psi_a(x,p)= W(\psi_a,\psi_a)(x,p)$, we have:
\begin{equation}
A \star W\psi_a= W\psi_a \star A= a W\psi_a~.
\label{eqWignerFunctions30}
\end{equation} 

\subsection{The Wigner-Moyal equation}

Let $\widehat{\rho} (t)$ denote the density matrix (regardless of whether the state is pure or mixed) of some quantum mechanical system at a given time $t$. The dynamics is governed by the von Neumann equation:
\begin{equation}
i \hbar \frac{\partial}{\partial t} \widehat{\rho}  (t) = \left[\widehat{H}, \widehat{\rho} (t) \right]~, 
\label{eqWignerFunctions31}
\end{equation}  
 where $\widehat{H}$ is the Hamiltonian, which we assume to be Hermitian and time-independent. Denoting by $W \rho (x,p,t)$ the Wigner function associated with $\widehat{\rho} (t)$ and by $H(x,p)$ the Weyl symbol of $\widehat{H}$, we obtain from (\ref{eqWignerFunctions24},\ref{eqWignerFunctions31}) :
\begin{equation}
\frac{\partial }{\partial t} W \rho (x,p,t)= \left[H(x,p), W \rho(x,p,t) \right]_M~.
\label{eqWignerFunctions32}
\end{equation}  
If the hamiltonian is of the form $H(x,p)= \frac{p^{2}}{2m} + V(x)$, then (\ref{eqWignerFunctions32}) yields:

\begin{equation}
\begin{array}{c}
\frac{\partial }{\partial t} W \rho (x,p,t)= -
 \frac{p}{m} \cdot \nabla_x W \rho (x,p,t)- \frac{1}{\hbar (\pi \hbar)^d} \int \mathcal{V}(x,p^{\prime}) W \rho (x,p+p^{\prime},t) dp^{\prime}=\\
 \\
 = \frac{p}{m} \cdot \nabla_x W \rho (x,p,t)+ \nabla_x V(x)  \cdot \nabla_p W \rho (x,p,t) + \mathcal{O} (\hbar^2) ~,
\end{array}
\label{eqWignerFunctions33}
\end{equation}  
where
\begin{equation}
\mathcal{V}(x,p^{\prime})= \int \sin \left(\frac{x^{\prime} p^{\prime}}{\hbar} \right) 
\left[V\left(x+ \frac{x^{\prime}}{2}\right) -V\left(x- \frac{x^{\prime}}{2}\right) \right]d x^{\prime}~.
\label{eqWignerFunctions34}
\end{equation} 

\section{Quantum devices}

When modelling a quantum device, one usually considers that the device is contained in an interval, say $I=\left[0, L\right]$, where the wavefunction is subjected to some potential $V_1(x)$, and that outside of the interval some incoming wave is under the action of some other potential $V_2(x)$.  

The potential on $\mathbb{R}$ can thus be written as:
\begin{equation}
V(x)= \theta_I (x) V_2 (x)+ \theta_{I^C}(x) V_1 (x)~,
\label{eqProfiles9}
\end{equation}
where $\theta_I (x)$ and $\theta_{I^C}(x)$ are the characteristic functions of the interval $I=\left[0,L\right]$ and of its complement, respectively:
\begin{equation}
\theta_I (x)= H(x)H(L-x)
 =\left\{
\begin{array}{l l}
1, & \text{if } x \in I\\
& \\
0, & \text{if } x \notin I
\end{array}
\right.
\label{eqProfiles2A}
\end{equation}
\begin{equation}
\theta_{I^C} (x)= \left[H(-x)+H(x-L)\right]=\left\{
\begin{array}{l l}
1, & \text{if } x \notin I\\
& \\
0, & \text{if } x \in I
\end{array}
\right.
\label{eqProfiles2B}
\end{equation}
Here $H(x)$ is Heaviside's step function.

Similarly, the solution $\psi$ of the time independent Schr\"odinger equation
\begin{equation}
- \frac{\hbar^2}{2m} \psi^{\prime \prime} (x) + V(x) \psi(x) = E \psi(x)~,
\label{eqProfiles2C}
\end{equation}
can be written in the form 
\begin{equation}
 \psi(x)=  \theta_{I^C}(x) \chi (x)+  \theta_I (x)  \phi(x)~,
\label{eqProfiles5}
\end{equation}
where $\phi(x)$ is the bulk function, which is a solution of
\begin{equation}
-\frac{\hbar^2}{2m} \phi^{\prime \prime} (x)+ V_1(x) \phi (x)= E \phi(x)~,
\label{eqProfiles6}
\end{equation}
whereas $\chi$ is the wavefunction outside of the device, satisfying
\begin{equation}
-\frac{\hbar^2}{2m} \chi^{\prime \prime} (x)+ V_2(x) \chi (x)= E \chi(x)~.
\label{eqProfiles7}
\end{equation}
We may also have to impose some boundary conditions, such as continuity of the wave function or its derivative, but we ignore that issue for now.

By the bilinearity of the Wigner transform, we have:
\begin{equation}
\begin{array}{c}
W \psi(x,p)=W\left(\theta_I \phi\right) (x,p)+W\left(\theta_I \phi,\theta_{I^C} \chi \right) (x,p)+\\
\\
+W\left(\theta_{I^C}\chi, \theta_I \phi\right)(x,p)+W\left(\theta_{I^C} \chi\right)(x,p)~.
\end{array}
\label{eqProfiles3A}
\end{equation}
From the support property (cf. section \ref{SubsectionSupportProperty}), the function $W\left(\theta_I \phi\right)$ is supported on the infinite strip $I \times \mathbb{R}$, while $W\left(\theta_{I^C} \chi\right)$ is supported on $I^C \times \mathbb{R}$. However, the interference term $W\left(\theta_I \phi,\theta_{I^C} \chi \right)+W\left(\theta_{I^C}\chi, \theta_I \phi\right)$ may have support on the entire phase space $\mathbb{R}^2$. 

Now suppose that we want to solve the stargenvalue equation associated to the Schr\"odinger equation (\ref{eqProfiles2C}) or instead the dynamical Wigner-Moyal equation. Then one often
wants to consider only the part of the wavefunction (or of the Wigner function) inside of the device. The reason is, of course, that from a numerical point of view, solving the equations is simpler in a bounded interval than on the whole real line. The price to pay is that the system then becomes an open system and the influence of the exterior of the device has to be accounted for with some suitable boundary conditions. 

Let us denote by $\widehat{\mathcal{P}}_{I}$ the operator which projects phase space functions $F(x,p)$ onto the infinite strip $I \times \mathbb{R}=\left\{(x,p) \in \mathbb{R}^2:~0\leq x \leq L \right\}$:
\begin{equation}
\widehat{\mathcal{P}}_{I} \, F = \theta_I\, F~,
\label{eqBoundaries36}
\end{equation}

We then have two alternative descriptions of the bulk system:
\begin{itemize}
\item {\bf Method 1} We perform the projection onto the interval within the Schr\"odinger formalism adding, for instance, Lent-Kirkner boundary conditions \cite{Lent}. We obtain the function 
$ \theta_I (x) \phi (x)$, where $\phi$ is a solution of (\ref{eqProfiles6}) satisfying the appropriate boundary conditions. Acting on this function with the Wigner transform yields:
\begin{equation}
F_{Bulk}(x,p)=W \left(\theta_I  \phi \right) (x,p)~.
\label{eqMethod11}
\end{equation}
The resulting function is a Wigner function with support on the infinite strip $I \times \mathbb{R}$.  

\vspace{0.2 cm}
\item {\bf Method 2} Alternatively, we may apply directly the Wigner transform to the entire function $\psi(x)$, $x \in \mathbb{R}$, and obtain (\ref{eqProfiles3A}). We then project the resulting function onto the infinite strip $I \times \mathbb{R}$, with the help of operator (\ref{eqBoundaries36}). As a result, we obtain:
\begin{equation}
\begin{array}{c}
F_{Bulk}(x,p)= \left(\widehat{\mathcal{P}}_I W \psi \right)(x,p)= W\left(\theta_I \phi\right) (x,p)+\\
\\
+ \left(\widehat{\mathcal{P}}_IW\left( \theta_I \phi,\theta_{I^C} \chi \right) \right) (x,p)
+\left(\widehat{\mathcal{P}}_I W\left(\theta_{I^C}\chi, \theta_I \phi\right)\right)(x,p)~.
\end{array}
\label{eqProfiles4A}
\end{equation}
\end{itemize}
We will discuss the virtues and shortcomings of these two methods in the next two sections.

\section{On the first Method}

In this section, we wish to concentrate on the first method, and try to illustrate with a concrete example the kind of difficulties that one encounters when considering the Wigner formulation of a system described by a confined wave function.

To be concrete, we consider a free particle, with wave function $\psi(x)$, in a one-dimensional box of size $L$, with infinite potential barriers located at $x=0$ and $x=L$. Hence, we have that 
\begin{equation}
\psi(x)=0~, \text{ for } x \notin \left(0,L\right)~.  
\label{eqBoundaries1}
\end{equation}
The time independent Schr\"odinger equation for $x \in \left(0,L \right)$ reads:
\begin{equation}
- \frac{\hbar^2}{2m} \psi^{\prime \prime} (x) = E \psi(x)~,
\label{eqBoundaries1}
\end{equation}
where $E$ is the energy. If we impose Dirichlet boundary conditions (to ensure hermiticity of the Hamiltonian)
\begin{equation}
\psi\left(0 \right)=\psi(L)=0~,
\label{eqBoundaries2}
\end{equation}
the energy becomes quantized:
\begin{equation}
E_n= \frac{ n^2 \pi^2 \hbar^2}{2mL^2}~, n=1,2,3, \cdots,
\label{eqBoundaries3}
\end{equation}
with associated eigenstates:
\begin{equation}
\psi_n(x) = \sqrt{\frac{2}{L}} \sin \left(\frac{n \pi x}{L} \right)~.
\label{eqBoundaries4}
\end{equation}
Let us now consider the Weyl-Wigner formulation of a slightly more general case of a system confined to the interval $\left[a,b \right]$ (we will set $a=0$, $b=L$ later) with wave function $\psi$.

From the discussion in section \ref{SubsectionSupportProperty}, we know that the Wigner function is also confined to the interval $[a,b]$. Moreover, we have proven in \cite{Dias1} that it can be written in the form:
\begin{equation}
W \psi(x,p)= \Theta_1 (x) F_1 (x,p)+ \Theta_2 (x) F_2 (x,p)~,
\label{eqBoundaries5}
\end{equation}
where
\begin{equation}
\Theta_1(x)= \left\{
\begin{array}{l l}
1, & \text{if } a<x \leq x_0 \\
0, & \text{otherwise}
\end{array}
\right. \hspace{1 cm} 
\Theta_2(x)= \left\{
\begin{array}{l l}
1, & \text{if } x_0<x <b \\
0, & \text{otherwise}
\end{array}
\right.~,
\label{eqBoundaries6}
\end{equation}
with $x_0= \frac{a+b}{2}$, and
\begin{equation}
F_1(x,p)= \frac{1}{\pi \hbar} \int_{a-x}^{x-a}e^{-2i py / \hbar} \psi(x+y) \psi^{\ast} (x-y) dy~,
\label{eqBoundaries7}
\end{equation}
\begin{equation}
F_2(x,p)= \frac{1}{\pi \hbar} \int_{x-b}^{b-x}e^{-2i py / \hbar} \psi(x+y) \psi^{\ast} (x-y) dy~,
\label{eqBoundaries8}
\end{equation}
For more details, see \cite{Dias1}.

\subsection{Boundary conditions}

First of all notice that, regardless of the boundary conditions satisfied by $\psi$ at $x=a$ or $x=b$, the Wigner function will always satisfy Dirichlet boundary conditions:
\begin{equation}
W \psi(a,p)=W \psi(b,p)=0~, \text{ for all } p \in \mathbb{R}~.
\label{eqBoundaries9}
\end{equation}
This can be seen directly from (\ref{eqBoundaries5}-\ref{eqBoundaries8}), but for all practical purposes, it is an immediate consequence of the fact that $W \psi$ is a uniformly continuous function on $\mathbb{R}^2$ and the fact that it is supported on $[a,b] \times \RE$. If one wants to find out the effect of the boundary conditions for $\psi$ on the Wigner function, one has to dig deeper. Indeed, it can be shown from (\ref{eqBoundaries5}-\ref{eqBoundaries8}), that, if $\psi$ satisfies \textit{Dirichlet} boundary conditions: 
\begin{equation}
\psi(a^+)=\psi(b^-)=0~,
\label{eqBoundaries10}
\end{equation}
then \cite{Dias1}:
\begin{equation}
\frac{\partial}{\partial x} W \psi(a^+,p)=\frac{\partial}{\partial x} W \psi(b^-,p)=\frac{\partial^2}{\partial x^2} W \psi(a^+,p)=\frac{\partial^2}{\partial x^2} W \psi(b^-,p)
=0~, 
\label{eqBoundaries11}
\end{equation}
for all $p \in \mathbb{R}$. However, in general, we have $\frac{\partial^3}{\partial x^3} W \psi(a^+,p) \neq 0$ and $\frac{\partial^3}{\partial x^3} W \psi(b^-,p) \neq 0$.

Alternatively, if $\psi$ is subjected to Neumann conditions,
\begin{equation}
\psi^{\prime}(a^+)=\psi^{\prime}(b^-)=0~,
\label{eqBoundaries12}
\end{equation}
then \cite{Dias1}:
\begin{equation}
\frac{\partial^2}{\partial x^2} W \psi(a^+,p)=\frac{\partial^2}{\partial x^2} W \psi(b^-,p)
=0~, 
\label{eqBoundaries13}
\end{equation}
for all $p \in \mathbb{R}$, but in general, we have $\frac{\partial}{\partial x} W \psi(a^+,p) \neq 0$ and $\frac{\partial}{\partial x} W \psi(b^-,p) \neq 0$.

In this context it is worth to remark that the boundary conditions themselves may have an effect in the bulk. For instance, consider the middle point of the interval, $x_0= \frac{a+b}{2}$. Albeit everywhere continuous, the Wigner function need not be differentiable. But, if $\psi$ satisfies the Dirichlet conditions (\ref{eqBoundaries10}), then:
\begin{equation}
\frac{\partial}{\partial x} W \psi(x_0^+,p)=\frac{\partial}{\partial x} W \psi(x_0^-,p)~,
\label{eqBoundaries14}
\end{equation}
for all $p \in \mathbb{R}$. However, if $\psi$ satisfies Neumann boundary conditions (\ref{eqBoundaries12}), then (\ref{eqBoundaries14}) does not hold in general.

\subsection{The trouble with the stargenvalue equation}

Let us go back to the example of the particle in the box (\ref{eqBoundaries1}-\ref{eqBoundaries4}). The Wigner function associated with the eigenstate $\psi_n(x)$ is:
\begin{equation}
\begin{array}{c}
W \psi_n (x,p)= \frac{1}{2 \pi \hbar L} \left\{\frac{\sin \left[2 \left(L/2-\left|x-L/2\right| \right) \left(n\pi /L-p/ \hbar\right) \right]}{n \pi/ L- p/\hbar}+ \right.\\
\\
\left. + \frac{\sin \left[2 \left(L/2-\left|x-L/2\right| \right) \left(n\pi /L+p/ \hbar\right) \right]}{n \pi/ L+ p/\hbar} \right\} + \frac{1}{\pi p L}  \cos \left(2 n \pi x/L\right) \sin\left[\frac{2p}{\hbar} \left(\left|x-L/2\right|- L/2 \right) \right]~,
\end{array}
\label{eqBoundaries15}
\end{equation}
for $0 < x < L$, and vanishes identically otherwise. The value at the critical points should be understood in sense of the limit, to ensure a proper continuous function. For instance:
\begin{equation}
\begin{array}{c}
W \psi_n (x,n\pi \hbar /L)=\lim_{p \to n\pi \hbar /L}W \psi_n (x,p)=\\
\\
=\frac{L/2-\left|x-L/2\right|}{\pi \hbar L} + \frac{1}{4 \pi^2 n \hbar} \sin \left[\frac{4n \pi}{L}\left(L/2 - \left|x-L/2 \right| \right) \right] + \\
\\
+\frac{1}{n \pi^2 \hbar} \cos \left(2 n \pi x/L\right) \sin \left(\frac{2n \pi}{L} \left(\left|x-L/2 \right|- L/2 \right) \right]~.
\end{array}
\label{eqBoundaries16}
\end{equation}
The stargenvalue equation corresponding to the eigenvalue equation (\ref{eqBoundaries1}) reads:
\begin{equation}
\begin{array}{c}
\frac{p^2}{2m} \star W \psi_n (x,p)= E_n W \psi_n (x,p) \\
\\
\Leftrightarrow \frac{p^2}{2m} W \psi_n (x,p)- \frac{i \hbar p}{2m} \frac{\partial}{\partial x} W \psi_n (x,p) - \frac{\hbar^2}{8m} \frac{\partial^2}{\partial x^2} W \psi_n (x,p)= E_n W \psi_n (x,p)~.
\end{array}
\label{eqBoundaries17}
\end{equation}
Since $W \psi_n (x,p)$ is a real function, the imaginary part of the previous equation implies:
\begin{equation}
p \frac{\partial}{\partial x} W \psi_n (x,p)=0 ~.
\label{eqBoundaries18}
\end{equation}
Thus, for all $p\neq 0$ and all $x \in \mathbb{R}$, such that $W \psi_n$ is differentiable, we have that $ \frac{\partial}{\partial x} W \psi_n (x,p)=0$. This shows that (\ref{eqBoundaries15}) cannot be a solution of the stargenvalue equation (\ref{eqBoundaries17}).

Going back to generic function $\psi$ confined to the interval, let us compute $\frac{p^2}{2m} \star W \psi$. We obtain (see (\ref{eqBoundaries5}-\ref{eqBoundaries8})):
\begin{equation}
\left\{
\begin{array}{l l}
\frac{p^2}{2m}\star (\Theta_1(x) F_1(x,p)) =E (\Theta_1(x)F_1(x,p)) + \mathcal{B}_1(x,p)~, & \text{for } a<x<x_0\\
& \\
\frac{p^2}{2m}\star (\Theta_2(x) F_2(x,p))=E(\Theta_2(x)F_2(x,p)) + \mathcal{B}_2(x,p)~, & \text{for } x_0<x<b
\end{array}
\right.
\label{eqBoundaries19}
\end{equation}
The  extra \textit{boundary} term $\mathcal{B}_1(x,p)$ for $a<x<x_0$ is given by:
\begin{equation}
\begin{array}{c}
\mathcal{B}_1(x,p)=- \frac{\hbar}{2 \pi m} e^{- \frac{2ip}{\hbar}(a-x)} \left\{\frac{2ip}{\hbar} \psi^{\ast} (2x-a) \psi(a^+)+ \right.\\
\\
\left. + \psi^{\prime \ast} (2x-a) \psi(a^+)+ \psi^{\ast} (2x-a)\psi^{\prime} (a^+) \right\}
\end{array}
\label{eqBoundaries20}
\end{equation}
In particular, for Dirichlet boundary conditions ($\psi(a^+)=0$), we have:
\begin{equation}
\mathcal{B}_1^D(x,p)=- \frac{\hbar}{2 \pi m} e^{- \frac{2ip}{\hbar}(a-x)}\psi^{\ast} (2x-a)\psi^{\prime} (a^+)~.
\label{eqBoundaries20}
\end{equation}
From this expression, we realize that the presence of the boundary is felt well inside the bulk region $a<x<x_0$. This is a manifestation of the strongly non-local character of the Weyl-Wigner quantization. 

Let us next remark that the boundary term can be expressed in terms of the Wigner function $F_1$. For $\epsilon>0$, consider the following integral
\begin{equation}
\Lambda_{\epsilon} (x,p)= \int_{\mathbb{R}}k e^{ik(x-a)} F_1\left(x+ \epsilon,p-\frac{\hbar}{2}k \right) dk~.
\label{eqBoundaries21}
\end{equation}
If we plug (\ref{eqBoundaries7}) in (\ref{eqBoundaries21}), we obtain:
\begin{equation}
\begin{array}{c}
\Lambda_{\epsilon} (x,p)= \frac{2i}{\hbar}e^{-\frac{2ip}{\hbar} (a-x)} \left\{-\frac{2ip}{\hbar}\psi^{\ast} (2x-a+ \epsilon) \psi(a+ \epsilon)\right.\\
\\
\left.- \psi^{\prime \ast} (2x-a+\epsilon)\psi(a+\epsilon)+ \psi^{\ast}(2x-a+\epsilon) \psi^{\prime} (a+\epsilon) \right\}
\end{array}
\label{eqBoundaries22}
\end{equation}
For Dirichlet boundary conditions, we obtain, by taking the limit $\epsilon\to 0^+$:
\begin{equation}
\mathcal{B}_1^D(x,p)= \lim_{\epsilon \to 0^+} \frac{i \hbar^2}{4 \pi m}\Lambda_{\epsilon}^D (x,p)~.
\label{eqBoundaries23}
\end{equation}
To make the boundary term a bit more intelligible, we remark that, using the formula (\ref{eqWignerFunctions21}) for the star-product, we can show that:
\begin{equation}
\Lambda_{\epsilon} (x,p)=-2 i \pi \delta^{\prime} (x-a) \star (\Theta_1(x+\epsilon) F_1 (x+\epsilon,p))~.
\label{eqBoundaries24}
\end{equation}
Let us define the operator
\begin{equation}
\delta_+(x) \phi(x) = \lim_{\epsilon \to 0^+} \delta(x) \phi(x+ \epsilon)~.
\label{eqBoundaries25}
\end{equation}
Notice that if $\phi$ is continuous at $x=0$, then there is no difference between the action of $\delta_+(x)$ and $\delta(x)$ on $\phi$:
\begin{equation}
\delta_+(x) \phi(x) =\delta(x) \phi(x)=\delta(x) \phi(0)~.
\label{eqBoundaries26}
\end{equation}
However, if $\phi$ is discontinuous at $x=0$ with finite lateral limit $\phi(0^+)$, then we have that $\delta(x) \phi(x)$ is ill-defined, while:
\begin{equation}
\delta_+(x) \phi(x) =\delta(x) \phi(0^+)~.
\label{eqBoundaries27}
\end{equation}
In a similar fashion, we have:
\begin{equation}
\delta_+^{\prime}(x) \phi(x) =\delta^{\prime}(x) \phi(0^+)-\delta(x) \phi^{\prime}(0^+)~.
\label{eqBoundaries28}
\end{equation}

We may now rewrite (\ref{eqBoundaries24}) as:
\begin{equation}
\lim_{\epsilon \to 0^+} \Lambda_{\epsilon}^D (x,p)=-2 i \pi \delta_+^{\prime} (x-a) \star (\Theta_1(x) F_1 (x,p))~.
\label{eqBoundaries29}
\end{equation}
Altogether, from (\ref{eqBoundaries19},\ref{eqBoundaries23},\ref{eqBoundaries29}), conclude that:
\begin{equation}
\left(\frac{p^2}{2m}- \frac{\hbar^2}{2m} \delta_+^{\prime} (x-a) \right) \star (\Theta_1(x) F_1(x,p))=E (\Theta_1(x) F_1(x,p)) ~,
\label{eqBoundaries30}
\end{equation}
for $a\leq x \leq x_0$, and assuming Dirichlet boundary conditions.

Analogously, we can show that
\begin{equation}
\left(\frac{p^2}{2m}+ \frac{\hbar^2}{2m} \delta_-^{\prime} (x-b) \right) \star (\Theta_2(x)F_2(x,p))=E (\Theta_2(x)F_2(x,p)) ~,
\label{eqBoundaries30}
\end{equation}
for $x_0 \leq x \leq b$, and where $\delta_- (x)$ is defined by 
$\delta_-(x) \phi(x)= \lim_{\epsilon \to 0^+} \delta (x) \phi(x- \epsilon)= \delta(x) \phi(0^-)$.

Altogether, we obtain for $W \psi$ (cf.(\ref{eqBoundaries5})):
\begin{equation}
\left(\frac{p^2}{2m} - \frac{\hbar^2}{2m} \delta_+^{\prime} (x-a)+ \frac{\hbar^2}{2m} \delta_-^{\prime} (x-b)  \right)\star W \psi(x,p)= E W \psi(x,p)~.
\label{eqBoundaries31}
\end{equation}

Before we conclude this section, let us make a few brief remarks.

\begin{enumerate}
\item If we add a potential of the form $V=V(x)$, then for Dirichlet boundary conditions, we obtain:
\begin{equation}
\left(\frac{p^2}{2m}+V(x) - \frac{\hbar^2}{2m} \delta_+^{\prime} (x-a)+ \frac{\hbar^2}{2m} \delta_-^{\prime} (x-b)  \right)\star W \psi(x,p)= E W \psi(x,p)~.
\label{eqBoundaries32}
\end{equation}

\vspace{0.2 cm}
\item The operators $\delta_+$ and $\delta_-$ can be defined rigorously in terms of an appropriate product of distributions. For the technical details see \cite{Dias3}-\cite{Dias6}.
of 
\vspace{0.2 cm}
\item Kryukov and Walton proposed an alternative approach to the stargenvalue equation \cite{Kryukov1,Walton}, which does not require boundary potentials, but which leads to higher-order partial differential equations in the phase space. Moreover, the dynamics in that context becomes more intricate. The connection between the two approaches is discussed in \cite{Dias2}.

\end{enumerate}

\subsection{Can we trace the origin of the boundary potentials?}

At first sight, it seems awkward that there is no need to consider boundary potentials in the time independent Schr\"odinger equation, but when one considers its phase space counterpart they are unavoidable. With the benefit of hindsight, one just has to be cautious about the conditions for which the Weyl correspondence between the operator formulation of quantum mechanics and its phase space formulation is valid.

When we solve the Schr\"odinger equation, we assume $\psi=0$ outside of the interval $\left[a,b\right]$ and solve the differential equation inside the interval. In other words, suppose that $\phi(x)$ is a solution of the time independent Schr\"odinger equation
\begin{equation}
-\frac{\hbar^2}{2m} \phi^{\prime \prime} (x)+V(x) \phi(x) = E \phi(x)~,
\label{eqBoundaries33}
\end{equation}
for all $x \in \mathbb{R}$, and that $\phi$ satisfies the conditions
\begin{equation}
\phi(a^+)=\phi(b^-)=0~.
\label{eqBoundaries34}
\end{equation}
Then the confined solution is simply
\begin{equation}
\psi(x)= \left(\widehat{P}_{I} \phi \right)(x)=\left\{
\begin{array}{l l}
\phi(x), & \text{if } a \leq x \leq b\\
& \\
0, & \text{otherwise} 
\end{array}
\right.
~,
\label{eqBoundaries35}
\end{equation}
where $\widehat{P}_{I}$ is the projection operator onto the interval $I=\left[a,b\right]$.

Our previous analysis shows that:
\begin{equation}
W(\widehat{P}_{I}\phi) (x,p) \neq  \left(\widehat{\mathcal{P}}_{I}W\phi\right)(x,p)~, 
\label{eqBoundaries37}
\end{equation}
or operatorially:
\begin{equation}
\mathcal{W} \widehat{P}_{I} \neq \widehat{\mathcal{P}}_{I} \mathcal{W}~.
\label{eqBoundaries38}
\end{equation}
The reason is that Weyl quantization is intrinsically non-local, and so we can only apply the Weyl transform to the {\it global} Schr\"odinger formulation of the system. Hence the wave function has to be defined on the entire space. In other words (with the previous notation), we have to deal with the function $\psi$ as defined in (\ref{eqBoundaries35}) and not with $\phi$. 

To avoid unnecessary complications, let us assume only one boundary at $x=a$, and that the system is confined to the interval $x>a$. In other words, we assume
\begin{equation}
\psi(x)= H(x-a) \phi(x) = \left\{
\begin{array}{l l}
\phi(x), & \text{if } x \geq a \\
& \\
0, & \text{otherwise}
\end{array}
\right.
\label{eqBoundaries39}
\end{equation}
where $H(x)$ is Heaviside's step function, and $\phi$ is the (twice differentiable) solution of (\ref{eqBoundaries33}), satisfying $\phi(a)=0$. From (\ref{eqBoundaries39}) we conclude that $\psi$ has well defined lateral limits:
\begin{equation}
\psi(a^+)=\phi(a)=0~,  \hspace{1 cm} \psi^{\prime} (a^+)=\phi^{\prime} (a)~.
\label{eqBoundaries40}
\end{equation}
A straightforward calculation shows that:
\[
\psi^{\prime}(x)=\delta(x-a) \phi(x)+H(x-a) \phi^{\prime} (x)~,
\]
and thus:
\begin{equation}
\begin{array}{c}
\psi^{\prime \prime} (x)=\delta^{\prime} (x-a) \phi(x) + 2 \delta(x-a) \phi^{\prime}(x)+H(x-a) \phi^{\prime \prime}(x)\\
\\
\Leftrightarrow \psi^{\prime \prime} (x)=\delta^{\prime} (x-a) \phi(a) +  \delta(x-a) \phi^{\prime}(a)+H(x-a)\phi^{\prime \prime}(x)\\
\\
\Leftrightarrow \psi^{\prime \prime} (x)= \delta(x-a) \phi^{\prime}(a)+H(x-a) \phi^{\prime \prime}(x)
\end{array}
\label{eqBoundaries41}
\end{equation}
From (\ref{eqBoundaries33}) and (\ref{eqBoundaries41}), we obtain:
\begin{equation}
\begin{array}{c}
- \frac{\hbar^2}{2m} \psi^{\prime \prime} (x)+V(x) \psi(x)=- \frac{\hbar^2}{2m} \delta(x-a) \phi^{\prime} (a) + E H(x-a) \phi(x)\\
\\
\Leftrightarrow - \frac{\hbar^2}{2m} \psi^{\prime \prime} (x)+V(x) \psi(x)=- \frac{\hbar^2}{2m} \delta(x-a) \phi^{\prime} (a) + E  \psi(x)~.
\end{array}
\label{eqBoundaries42}
\end{equation}
On the other hand, from (\ref{eqBoundaries28}) and (\ref{eqBoundaries40}) we have:
\begin{equation}
 \delta(x-a) \phi^{\prime} (a)= \delta(x-a) \psi^{\prime} (a^+)=- \delta_+^{\prime}(x-a) \psi (x)~.
\label{eqBoundaries43}
\end{equation}
Altogether:
\begin{equation}
- \frac{\hbar^2}{2m} \psi^{\prime \prime} (x)+V(x) \psi(x)- \frac{\hbar^2}{2m} \delta_+^{\prime}(x-a) \psi (x) = E  \psi(x)~.
\label{eqBoundaries44}
\end{equation}
With another boundary at $x=b$ and Dirichlet condition $\phi(b)=0$, we would obtain:
\begin{equation}
- \frac{\hbar^2}{2m} \psi^{\prime \prime} (x)+V(x) \psi(x)- \frac{\hbar^2}{2m} \delta_+^{\prime}(x-a) \psi (x)+ \frac{\hbar^2}{2m} \delta_-^{\prime}(x-b) \psi (x) = E  \psi(x)~.
\label{eqBoundaries45}
\end{equation}
This shows that the {\it globally} defined Hamiltonian for a particle confined to the interval $I= \left[a,b \right]$ with Dirichlet boundary conditions is:
\begin{equation}
\widehat{H}= \frac{p^2}{2m}+V(x) - \frac{\hbar^2}{2m} \delta_+^{\prime}(x-a)+ \frac{\hbar^2}{2m} \delta_-^{\prime}(x-b)~.
\label{eqBoundaries46}
\end{equation}
This is why the corresponding stargenvalue equation is (\ref{eqBoundaries32}). 

The calculation (\ref{eqBoundaries41}) also shows that boundary conditions other than Dirichlet conditions will necessarily yield a different boundary potential.

\subsection{Dynamics for confined systems}

It is now obvious that, if a system is confined to an interval $I=\left[a,b \right]$ and the density matrix $\widehat{\rho} (t)$ with kernel $\rho(x,y)$ satisfies Dirichlet boundary conditions
\begin{equation}
\rho(a^+,y,t)=\rho(b^-,y,t)=\rho(x,a^+,t)=\rho(x,b^-,t)=0~, \text{ for all } x,y \in \mathbb{R}~,
\label{eqBoundaries47}
\end{equation}
and all $t$ in some interval, then the dynamics is dictated by the von Neumann equation (\ref{eqWignerFunctions31}) with Hamiltonian (\ref{eqBoundaries46}).

Consequently, the Wigner Moyal equation becomes:
\begin{equation}
\begin{array}{c}
\frac{\partial}{\partial t} W \rho(x,p,t)= \left[\frac{p^2}{2m}+V(x) - \frac{\hbar^2}{2m} \delta_+^{\prime}(x-a)+ \frac{\hbar^2}{2m} \delta_-^{\prime}(x-b), \rho(x,p,t) \right]_M=\\
\\
=- \frac{p}{m} \frac{\partial}{\partial x} W \rho(x,p,t) - \frac{1}{\pi \hbar^2} \int \mathcal{V}(x,p^{\prime}) W \rho(x,p+p^{\prime},t) d p^{\prime}~, 
\end{array}
\label{eqBoundaries48}
\end{equation}
where this time:
\begin{equation}
\begin{array}{c}
\mathcal{V}(x,p^{\prime})= \int \sin\left(\frac{x^{\prime}p^{\prime}}{\hbar} \right) \left\{V \left(x+ \frac{x^{\prime}}{2} \right)-\frac{\hbar^2}{2m}\delta_+^{\prime} \left(x+ \frac{x^{\prime}}{2} -a\right)\right.\\
\\
 +\frac{\hbar^2}{2m} \delta_-^{\prime} \left(x+ \frac{x^{\prime}}{2} +b\right)- V\left(x- \frac{x^{\prime}}{2} \right)+\\
 \\
 \left. + \frac{\hbar^2}{2m} \delta_+^{\prime}\left(x- \frac{x^{\prime}}{2} -a\right)-\frac{\hbar^2}{2m}\delta_-^{\prime}\left(x- \frac{x^{\prime}}{2} +b \right) \right\}d x^{\prime}
\end{array}
\label{eqBoundaries49}
\end{equation}

Thus, if we decide to follow Method 1, the projection $\widehat{P}_I \psi$ of the wavefunction $\psi$ onto the interval $I$, leads inevitably to the appearance of boundary potentials on top of the rather intricate additional boundary conditions (e.g. Lent-Kirkner boundary conditions).

\subsection{Numerical solution of the confined eigenvalue problem}

We now discuss the numerical implementation of the distributional boundary potentials. We will do this in the context of the energy eigenvalue equation for a localized one-dimensional quantum particle. We will use the Sch\"odinger representation since it yields an ordinary differential equation, while in the case of the Wigner formulation, we have to deal with an integral-differential equation. This latter case was considered in [section VII, \cite{Dias1}] where we studied the numerical formulation of the $*$-genvalue equation with distributional boundary potentials. We used it to model a confined quantum particle, and determined its Wigner trajectories. We also compared the numerical results with the exact solutions. The interested reader should refer to \cite{Dias1} for a detailed discussion. 

In this section we are interested in solving the eigenvalue equation     
\begin{equation}
- \frac{\hbar^2}{2m} \psi^{\prime \prime} (x) - \frac{\hbar^2}{2m} \delta_+^{\prime}(x+1) \psi (x)+ \frac{\hbar^2}{2m} \delta_-^{\prime}(x-1) \psi (x) = E  \psi(x)
\label{eqBoundaries45-1}
\end{equation}
which is obtained from (\ref{eqBoundaries45}) by setting $V=0,\, b=-a=1$; for Dirichlet boundary conditions. In order to model the distributional terms $\delta_{\pm}^{\prime}(x\pm 1) \psi (x) = \delta^{\prime}(x\pm 1) \psi (x^{\pm})$ we approximate the Dirac delta function by a Gaussian:
\begin{equation}
\delta_\epsilon(x)= \frac{1}{\pi^{1/2} \epsilon} \exp \left[- \left(\frac{x}{\epsilon}\right)^2 \right]
\label{Gaussian}	
\end{equation}
and the "shifted" $\psi$ by $\psi(x^\pm) \simeq \psi(x\pm 2\epsilon)$. The exact equation is obtained by taking the limit $\epsilon \to 0^+$, in which case the boundary term yields the product of $\delta'(x\pm 1)$ by the left (or right) limit of $\psi$ at $x=\pm 1$ (notice that $\psi$ or its derivatives may be discontinuous at $x= \pm 1$).

We thus want to solve the following eigenvalue equation numerically:
\begin{equation}
- \frac{\hbar^2}{2m} \psi_\epsilon^{\prime \prime} (x) - \frac{\hbar^2}{2m} \delta_\epsilon^{\prime}(x+1) \psi_\epsilon (x+2\epsilon)+ \frac{\hbar^2}{2m} \delta_\epsilon^{\prime}(x-1) \psi_\epsilon (x-2\epsilon) = E  \psi_\epsilon(x)
\label{numericalEq}	
\end{equation}
for Dirichlet boundary conditions $\psi(-1^+)=\psi(1^-)=0$. We implement these on $\psi_\epsilon$ as:
\begin{equation}
\psi_\epsilon(-1+2\epsilon)=\psi_\epsilon(1-2\epsilon) =\phi_1(1-2\epsilon)
\label{BC}
\end{equation}
where $\phi_1$ is the ground state of the free particle Hamiltonian satisfying Dirichlet boundary conditions at $x=\pm 1$.

Figure 1 displays the graph of $\phi_1$, solution of (\ref{eqBoundaries33}) for $V=0$. We set $E=E_1=\pi^2/8$ (the lowest eigenvalue for Dirichlet boundary conditions at $x=\pm 1$) and $\hbar=m=1$.

\begin{figure}[h]
\center\includegraphics [scale=0.5] {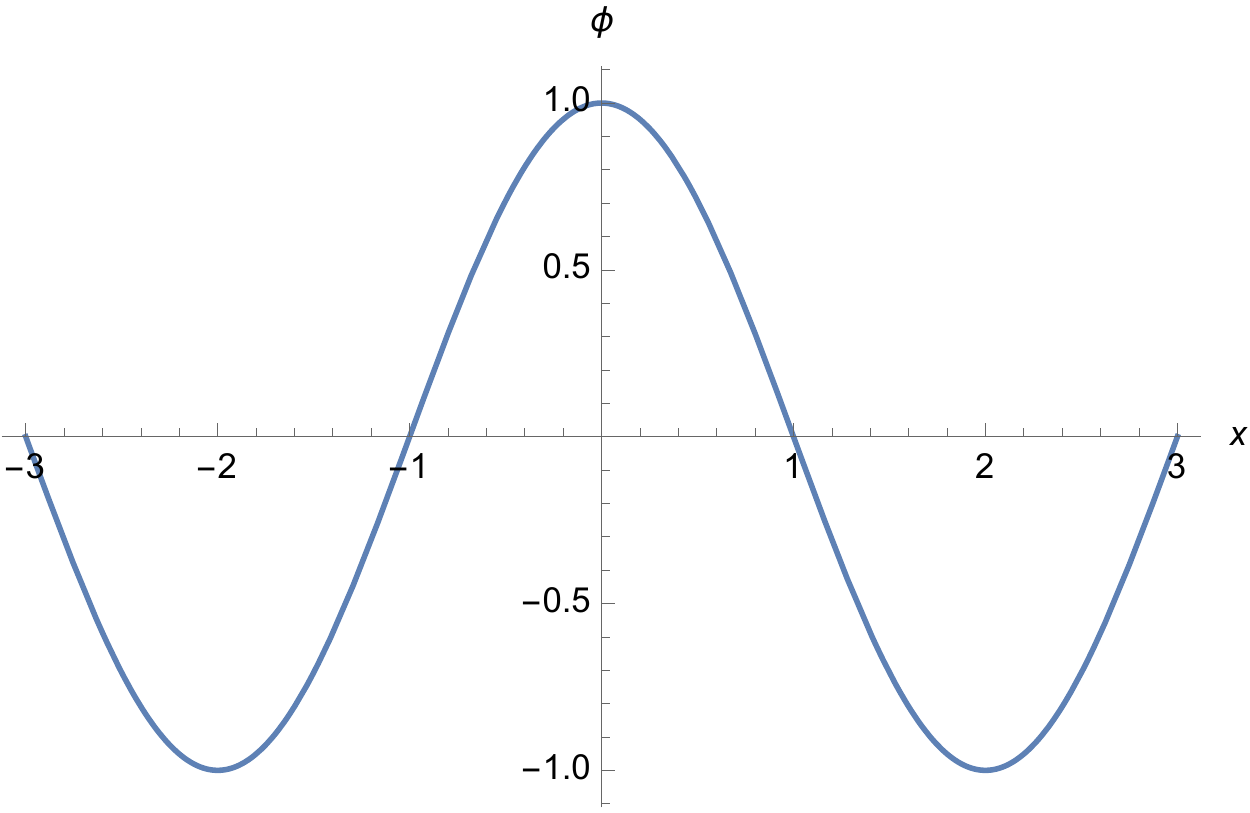}
\end{figure}
\vspace{4.9cm}
\footnotesize {Fig.1: Ground state of the free particle Hamiltonian satisfying Dirichlet boundary conditions at $x=\pm 1$ and for $\hbar=m=1$.} 
\vspace{0.3cm}

\normalsize

Figures 2 and 3 display the numerical solutions of (\ref{numericalEq},\ref{BC}) for $\epsilon =0.01$ and $\epsilon =0.0025$, respectively. 

\begin{figure}[h]
\center\includegraphics [scale=0.5] {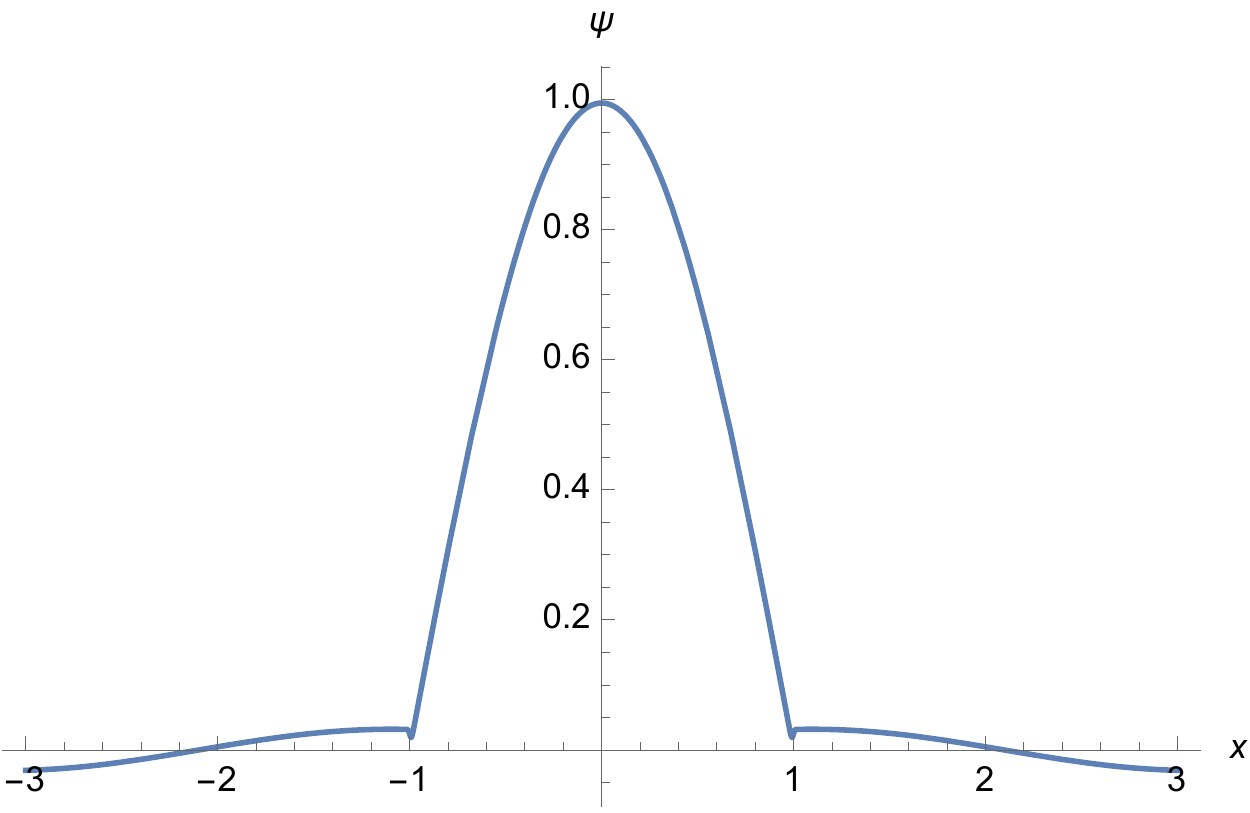}
\end{figure}

\footnotesize {Fig.2: Numerical solution of eq.(\ref{numericalEq}) with boundary conditions (\ref{BC}): $\epsilon=0.01$, $\hbar=m=1$ and $E=\pi^2/8$.} 
\vspace{0.3cm}

\normalsize

\begin{figure}[h]
\center\includegraphics [scale=0.5] {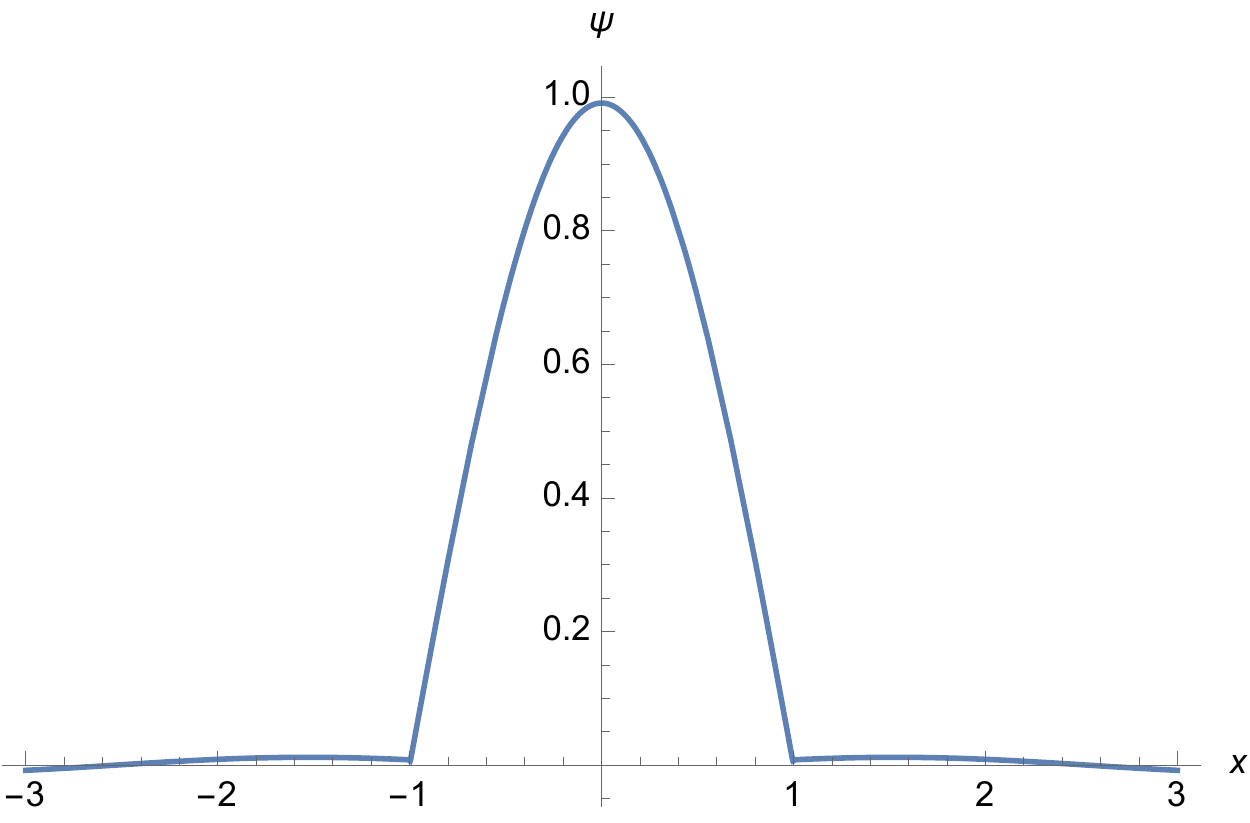}
\end{figure}

\footnotesize {Fig.3: Numerical solution of eq.(\ref{numericalEq}) with boundary conditions (\ref{BC}): $\epsilon=0.0025$, $\hbar=m=1$ and $E=\pi^2/8$.} 
\vspace{0.3cm}

\normalsize

Further discussion of these results is given in \cite{Dias1}.

\section{On the second Method}

We now follow the second Method. We thus write (\ref{eqProfiles5}) for the wavefunction. The associated Wigner transform is then given by (\ref{eqProfiles3A}) and the projection of the Wigner function onto the infinite strip is given by (\ref{eqProfiles4A}).

In this section, we discuss some of the difficulties that we have to deal with, if we choose the second Method.

\subsection{Positivity}\label{Positivity}

The first problem that we encounter is that $F_{Bulk}(x,p,t)$ is {\bf never} a Wigner function. The reason is that it fails to satisfy the positivity condition (\ref{eqWignerFunctions5B}).

In the ensuing analysis we will, by abuse of notation, write simply $\phi$ instead of $\theta_I\phi$ and $\chi$ instead of $\theta_{I^C} \chi$. We just have to bear in mind that $\phi$ has support in the interval $I=\left[0,L \right]$ whereas $\chi$ is supported on its complement $I^C$. 

Next, consider the function
\begin{equation}
\psi(x) =\left\{
\begin{array}{l l}
 \phi (x), & \text{if } 0 <x \leq L \\
&\\
- N \chi (x) , & \text{if } L <x \leq 2L \\
&\\
0, & \text{otherwise}
\end{array}
\right.
\label{eqPositivity1}
\end{equation}
The positive constant $N$ is required to satisfy:
\begin{equation}
N > \frac{ \| \phi\|_{L^2 (\mathbb{R})}^2}{2\int_0^L dx \int_L^{2L-x}dy |\phi(x)|^2  |\chi (y)|^2  }~.
\label{eqPositivity2}
\end{equation}
We have:
\begin{equation}
\begin{array}{c}
\widehat{\mathcal{P}}_I W \left(\chi, \phi \right)(x,p)=\\
\\
=H(x)H(L-x) \frac{1}{\pi \hbar} \int_{- \infty}^{+ \infty} \chi (x+y) \phi^{\ast} (x-y) e^{- \frac{2i}{\hbar} py} dy~.
\end{array}
\label{eqPositivity3}
\end{equation}
Then, we have:
\begin{equation}
\begin{array}{c}
 \int_{- \infty}^{+ \infty} dx  \int_{- \infty}^{+ \infty} dp \widehat{\mathcal{P}}_I W \left(\chi, \phi \right)(x,p) W \psi (x,p)=\\
 \\
=  \int_{0}^{L} dx  \int_{- \infty}^{+ \infty} dp W \left(\chi, \phi \right)(x,p) W \psi (x,p)=\\
\\
=\frac{1}{(\pi \hbar)^2}  \int_{0}^{L} dx  \int_{- \infty}^{+ \infty} dp\int_{- \infty}^{+ \infty} dy \int_{- \infty}^{+ \infty} dz \chi(x+y) \phi^{\ast} (x-y)\times \\
\\
\times \psi(x+z) \psi^{\ast} (x-z) e^{-\frac{2i}{\hbar}p(y+z)}=\\
\\
= \frac{1}{\pi \hbar}  \int_{0}^{L} dx  \int_{- \infty}^{+ \infty}  dy  \chi(x+y)  \psi^{\ast} (x+y) \phi^{\ast} (x-y) \psi(x-y)
\end{array}
\label{eqPositivity4}
\end{equation}
If we perform the substitution
\begin{equation}
X=x-y~, \hspace{0.5 cm} Y=x+y~,
\label{eqPositivity5}
\end{equation}
and use (\ref{eqPositivity1}), we obtain:
\begin{equation}
\begin{array}{c}
 \int_{- \infty}^{+ \infty} dx  \int_{- \infty}^{+ \infty} dp \widehat{\mathcal{P}}_I W \left(\chi, \phi \right)(x,p) W \psi (x,p)=\\
 \\
 = \frac{1}{2\pi \hbar}  \int_{- \infty}^{+ \infty} dX \int_{-X}^{2L-X} dY \chi (Y) \psi^{\ast} (Y) \phi^{\ast} (X) \psi(X)=\\
 \\
 = - \frac{N}{2\pi \hbar}  \int_{0}^{L} dX |\phi(X)|^2 \int_{L}^{2L-X} dY |\chi (Y)|^2~.
 \end{array}
\label{eqPositivity6}
\end{equation}
We thus have from (\ref{eqWignerFunctions4Moyal},\ref{eqProfiles4A},\ref{eqPositivity2},\ref{eqPositivity6}):
\begin{equation}
\begin{array}{c}
 \int_{- \infty}^{+ \infty}dx  \int_{- \infty}^{+ \infty}dp F_{Bulk}(x,p) W \psi(x,p) =\\
 \\
 = \int_{- \infty}^{+ \infty}dx  \int_{- \infty}^{+ \infty}dp  \left\{W \phi(x,p) W \psi(x,p) + 2 W \psi(x,p)  \text{Re} \widehat{\mathcal{P}}_I W \left(\chi, \phi \right)(x,p)\right\}=\\
 \\
 =\frac{\left|\langle \phi| \psi\rangle \right|^2}{2 \pi \hbar} + 2 \text{Re}  \int_{- \infty}^{+ \infty}dx  \int_{- \infty}^{+ \infty}dp ~ W \psi(x,p) \widehat{\mathcal{P}}_I  W \left(\chi, \phi \right)(x,p)=\\
 \\
 =\frac{\| \phi\|_{L^2 (\mathbb{R})}^2}{2 \pi \hbar }-\frac{N}{\pi \hbar } \int_0^Ldx \int_L^{2L-x} dy |\phi(x)|^2  |\chi(y)|^2 <0~.
 \end{array}
\label{eqPositivity7}
\end{equation}
This shows that $F_{Bulk}(x,p)$ is never a Wigner function, except if it is completely confined: $\chi(x)=0$. 

\subsection{Dynamics of the bulk function}\label{BulkDynamics}

Suppose that the device is contained in the interval $I= \left[0,L\right]$ and that it is being subjected to some potential $V_1(x)$. The device is affected by an incoming wave (at $x=0,L$), where another potential $V_2(x)$ is in place. Thus we may assume that the global wavefunction is (\ref{eqProfiles5}) and the Schr\"odinger equation is given by (\ref{eqProfiles9},\ref{eqProfiles2C},\ref{eqProfiles5},\ref{eqProfiles6},\ref{eqProfiles7}). If we consider instead the time evolution of this system according to the von Neumann equation, then the Wigner-Moyal equation for the associated Wigner function is given by:
\begin{equation}
\frac{\partial}{\partial t} W \psi(x,p,t)= \left[\frac{p^2}{2m} +V(x) ,W \psi(x,p,t) \right]_M
\label{eqProfiles18}
\end{equation}
More generally, if we have a mixed state, whose Wigner function $W \rho(x,p,t)$ is a convex combination of functions satisfying (\ref{eqProfiles18}), then:
\begin{equation}
\begin{array}{c}
\frac{\partial}{\partial t} W \rho(x,p,t)= \left[\frac{p^2}{2m} +V(x),W \rho(x,p,t) \right]_M\\
\\
= - \frac{p}{m} \frac{\partial}{\partial x} W \rho(x,p,t) - \frac{1}{\pi \hbar^2} \int \mathcal{V} (x,p^{\prime})W \rho(x,p+p^{\prime},t) dp^{\prime}~,
\end{array}
\label{eqProfiles19}
\end{equation}
where
\begin{equation}
\begin{array}{c}
\mathcal{V}(x,p^{\prime})= \int  \sin \left(\frac{x^{\prime}p^{\prime}}{\hbar} \right) \left\{ V \left(x+\frac{x^{\prime}}{2} \right)-V\left(x-\frac{x^{\prime}}{2} \right))\right\} d x^{\prime}
\end{array}
\label{eqProfiles20}
\end{equation}
In this approach the equation for the dynamics of $F_{Bulk}(x,p,t)$ is simply given by (cf. (\ref{eqProfiles18})):
\begin{equation}
\frac{\partial}{\partial t} F_{Bulk}(x,p,t)=- \frac{p}{m} \frac{\partial}{\partial x} F_{Bulk} - \frac{1}{\pi \hbar^2} \int \mathcal{V}_{Bulk} (x,p^{\prime})F_{Bulk}(x,p+p^{\prime},t) dp^{\prime}~,
\label{eqProfiles21}
\end{equation}
where
\begin{equation}
\begin{array}{c}
\mathcal{V}_{Bulk} (x,p^{\prime})=\left(\widehat{\mathcal{P}}_I \mathcal{V}\right)(x,p^{\prime})=\\
\\
=H(x) H(L-x)\int  \sin \left(\frac{x^{\prime}p^{\prime}}{\hbar} \right) \left\{ V \left(x+\frac{x^{\prime}}{2} \right)-V\left(x-\frac{x^{\prime}}{2} \right)\right\} d x^{\prime}~.
\end{array}
\label{eqProfiles22}
\end{equation}
Several comments to this equation are in order:

\begin{enumerate}
\item It should be emphasized that the prevalent nonlocality of the Weyl-Wigner formulation takes its toll on the dynamical equation in the bulk. Although the potential $V_2(x)$ is present only outside of $I$, it nevertheless contributes to the bulk equation (\ref{eqProfiles21}) through the potential $V(x)$ appearing in (\ref{eqProfiles22}):
\begin{equation}
V(x)= \left[H(-x)+H(x-L)\right]V_2(x)+ H(x) H(L-x) V_1(x) ~.
\label{eqProfiles23}
\end{equation}

\vspace{0.2 cm}
\item In order to obtain a closed form formulation 
of (\ref{eqProfiles21}) we still have to provide appropriate boundary conditions at $x=0$ and $x=L$,
\begin{equation}
W \rho(x=0,p,t)=g_0(p,t)~, \hspace{1 cm}  W \rho(x=L,p,t)=g_L(p,t)~.
\label{eqProfiles24}
\end{equation}
These conditions should mimic the influence of  
$\widehat{\mathcal{P}}_{I^c} W\rho$ on the bulk. 
This was the procedure suggested by Frensley \cite{Frensley}.
We shall call $g_0(p,t)$ and $g_L(p,t)$ the \textit{profiles} of $W \rho$ at $x=0$ and $x=L$, respectively. In the next section, we will show that the choice of these profiles is delicate.

\vspace{0.2 cm}
\item Even if we obtain the exact dynamics of $F_{Bulk}(x,p,t)$, this is still not a proper Wigner function because we are neglecting parts of the Wigner function, which are needed to ensure the positivity. 

Nevertheless, it may be that Frensley's approach leads to a good local approximation to the exact Wigner function. In that case, it may be interesting to obtain the Wigner function which is \textit{closest} (in some appropriate sense) to this approximation, and to have a criterium for how good the approximation is. This will be the subject of section \ref{Approximation}.

\end{enumerate}

\section{Profiles of Wigner functions}

\subsection{Profile at one point}

Let $W \rho(x,p)$ be a some Wigner function on $\mathbb{R}^{2d}$. In what follows it is immaterial whether it is associated with a pure or a mixed state, and whether it is time dependent or stationary. The \textit{profile} of $W \rho(x,p)$ at a point $x=a$ is a function $g(p)$, such that
\begin{equation}
W \rho(a,p)=g(p)~, \text{ for all } p \in \mathbb{R}^d~.
\label{eqProfile1}
\end{equation}
The question is: what is required of an arbitrary function $g(p)$ to be a Wigner profile? The following theorem shows that very little is required. In the sequel, $\widetilde{g}$ denotes the Fourier transform:
\begin{equation}
\widetilde{g} (x)= \frac{1}{(2 \pi \hbar)^{d/2}}\int_{\mathbb{R}^d} g(p) e^{- \frac{i}{\hbar}x \cdot p} dp~.
\label{eqProfile2}
\end{equation}

\begin{theorem}\label{TheoremProfile1}
A measurable function $g: \mathbb{R}^d \to \mathbb{R}$ is a Wigner profile if and only if $g \in L^2 (\mathbb{R}^d)$ and $\widetilde{g} \in L^1 (\mathbb{R}^d)$ with
\begin{equation}
\|\widetilde{g}  \|_{L^1 (\mathbb{R}^d)}= \int_{\mathbb{R}^d}|\widetilde{g} (x)|dx \leq \left(\frac{2}{\pi \hbar}\right)^{d/2}~.
\label{eqProfile3}
\end{equation}
\end{theorem}

The full technical proof of this theorem will be given elsewhere \cite{Dias7}. Here, we will only sketch the proof.

Assume first that $g(p)$ is the profile of some Wigner function $W \rho(x,p)$ at $x=0$\footnote{By translation invariance of the Weyl transform, it is irrelevant whether we consider the profile at the origin or at another point $x=a$.}. Since $W \rho(x,p)$ is uniformly continuous on $\mathbb{R}^{2d}$ and belongs to $ L^2 (\mathbb{R}^{2d})$, we conclude that $g(p)=W \rho(0,p)$ belongs to $L^2 (\mathbb{R}^d)$.

Moreover, if $g(p)$ is the profile of some Wigner function
\begin{equation}
W \rho(x,p)=\sum_j p_j \left(\frac{1}{2 \pi \hbar}\right)^d \int_{\mathbb{R}^d} \psi_j\left(x+ \frac{y}{2} \right)  \psi_j^{\ast}\left(x- \frac{y}{2} \right)  e^{- \frac{i}{\hbar} p \cdot y} dy~,
\label{eqProfile4}
\end{equation}
then:
\begin{equation}
g(p)=W \rho(0,p)=\sum_j p_j \left(\frac{1}{2 \pi \hbar}\right)^d \int_{\mathbb{R}^d} \psi_j\left( \frac{y}{2} \right)  \psi_j^{\ast}\left( -\frac{y}{2} \right)  e^{- \frac{i}{\hbar} p \cdot y} dy~.
\label{eqProfile5}
\end{equation}
Consequently:
\begin{equation}
\widetilde{g}(x) = \left(\frac{1}{2\pi \hbar}\right)^{d/2} \sum_j p_j\psi_j\left( \frac{x}{2} \right)  \psi_j^{\ast}\left( -\frac{x}{2} \right)~.
\label{eqProfile6}
\end{equation}
From the Cauchy-Schwarz inequality:
\begin{equation}
\begin{array}{c}
\|\widetilde{g}\|_{L^1 (\mathbb{R}^d)}\leq \left(\frac{1}{2\pi \hbar}\right)^{d/2} \sum_j p_j \int_{\mathbb{R}^d} \left| \psi_j\left( \frac{x}{2} \right)  \psi_j^{\ast}\left( -\frac{x}{2} \right) \right| dx \\
\\
\leq \left(\frac{2}{\pi \hbar}\right)^{d/2} \sum_j p_j \|\psi_j\|_{L^2 (\mathbb{R}^d)}^2 = \left(\frac{2}{\pi \hbar}\right)^{d/2} ~.
\end{array}
\label{eqProfile6}
\end{equation}
Conversely, suppose that $g \in L^2 (\mathbb{R}^d)$ and that (\ref{eqProfile3}) holds. We want to prove that there exist Wigner functions such that $g(p)$ is their profile at $x=0$. 

Let
\begin{equation}
h(x)=(2 \pi \hbar)^{d/2} \widetilde{g}(-2 x)~,
\label{eqProfile7}
\end{equation}
and consider the polar decomposition of $h$:
\begin{equation}
h(x)= A(x) e^{\frac{i}{\hbar} B(x)}~.
\label{eqProfile8}
\end{equation}
Since $g(p)$ is a real function, we conclude that $\widetilde{g}^{\ast} (-x)= \widetilde{g}(x)$. This entails that
\begin{equation}
h^{\ast}(-x)=h(x)~.
\label{eqProfile9}
\end{equation}
In view of (\ref{eqProfile8}) and (\ref{eqProfile9}), it follows that $A(x)$ is an even function and $B(x)$ is an odd function:
\begin{equation}
A(-x)=A(x)~, \hspace{1 cm} B(-x)=-B(x)~.
\label{eqProfile10}
\end{equation}
Next, we define the following wave function:
\begin{equation}
\psi(x)= \sqrt{A(x)} \exp \left[O(x)+ \frac{i}{2\hbar} \left(B(x)+E(x) \right)\right]~,
\label{eqProfile11}
\end{equation}
where $O(x)$ is (for now) an arbitrary odd function and $E(x)$ an arbitrary even function. It can be shown \cite{Dias7} that the function $O(x)$ can be chosen in such a way that $\psi$ is normalized: $\|\psi\|_{L^2 (\mathbb{R}^d)}=1$. The proof is subtle, and requires the condition (\ref{eqProfile3}). We will not discuss that more technical part here.

The associated Wigner function is:
\begin{equation} 
\begin{array}{c}
W \psi(x,p)= \frac{1}{(2 \pi \hbar)^d} \int_{\mathbb{R}^d}\sqrt{A \left(x+ \frac{y}{2}\right)A \left(x- \frac{y}{2}\right) }\\
\\
 \exp \left[O\left(x+ \frac{y}{2}\right)+O\left(x- \frac{y}{2}\right) + \frac{i}{2\hbar} B \left(x+ \frac{y}{2}\right) - \frac{i}{2\hbar} B \left(x- \frac{y}{2}\right)+\right.\\
 \\
 \left.+ \frac{i}{2\hbar} E \left(x+ \frac{y}{2}\right)- \frac{i}{2\hbar} E \left(x+ \frac{y}{2}\right) - \frac{i}{\hbar} p \cdot y\right]   dy
\end{array}
\label{eqProfile12}
\end{equation}
Using the parity of $A,~B,~O,~E$, we obtain:
\begin{equation}
\begin{array}{c}
W \psi(0,p)= \frac{1}{(2 \pi \hbar)^d} \int_{\mathbb{R}^d}A \left(\frac{y}{2}\right)
 \exp \left[ \frac{i}{\hbar} B \left(\frac{y}{2}\right)  - \frac{i}{\hbar} p \cdot y\right]   dy=\\
 \\
 = \frac{1}{(2 \pi \hbar)^d} \int_{\mathbb{R}^d} h\left(\frac{y}{2}\right) e^{ - \frac{i}{\hbar} p \cdot y} dy =g(p)~,
\end{array}
\label{eqProfile13}
\end{equation}
which shows that $g(p)$ is the profile of $W \psi(x,p)$ at $x=0$.

The arbitrariness in the choice of the functions $O(x)$ and $E(x)$ allows us to do convex combinations and thus obtain Wigner functions associated with mixed states which have $g(p)$ as their profile at $x=0$. Thus, if we choose a set $\left\{O_j(x) \right\}$ of odd functions, a set $\left\{E_j(x) \right\}$ of even functions and a probability distribution $\left\{p_j \right\}$, we may define the Wigner function of a mixed state:
\begin{equation}
W \rho(x,p)=\sum_j p_j \left(\frac{1}{2 \pi \hbar}\right)^d \int_{\mathbb{R}^d} \psi_j\left( x+\frac{y}{2} \right)  \psi_j^{\ast}\left( x-\frac{y}{2} \right)  e^{- \frac{i}{\hbar} p \cdot y} dy~,
\label{eqProfile14}
\end{equation}
with 
\begin{equation}
\psi_j(x)= \sqrt{A(x)} \exp \left[O_j(x)+ \frac{i}{2\hbar} \left(B(x)+E_j(x) \right)\right]~.
\label{eqProfile15}
\end{equation}
And we have:
\begin{equation}
W \rho(0,p)= g(p)~.
\label{eqProfile16}
\end{equation}

\subsection{Profiles at two points}

Theorem \ref{TheoremProfile1} tells us that there is a lot of freedom in the choice of possible profiles. However, if we consider the profile at a second point, then we must impose an additional compatibility condition, at least for pure states.

\begin{theorem}\label{TheoremProfile2}
Two measurable functions $g_{\alpha}: \mathbb{R}^d \to \mathbb{R}$ $(\alpha=a,b)$ are the Wigner profiles of the same Wigner function $W \psi(x,p)$ at two different points $x=a$ and $x=b$, if and only if the following conditions are satisfied:

\begin{enumerate}
\item $g_{\alpha} \in L^2 (\mathbb{R}^d)$ and $\widetilde{g}_{\alpha} \in L^1 (\mathbb{R}^d)$, $\alpha =a,b$

\vspace{0.2 cm}
\item The following bound holds:
\begin{equation}
\|\widetilde{g}_{\alpha}\|_{L^1 (\mathbb{R}^d)} \leq \left(\frac{2}{\pi \hbar} \right)^{d/2}~,~~\alpha =a,b
\label{eqProfile17}
\end{equation}

\vspace{0.2 cm}
\item There exist two odd measurable functions $O_{\alpha}: \mathbb{R}^d \to \mathbb{R}$, two even measurable functions $E_{\alpha}: \mathbb{R}^d \to \mathbb{R}$, and two sign functions $sgn_{\alpha}: \mathbb{R}^d \to \left\{-1,+1 \right\}$, $\alpha=a,b$, such that:
\begin{equation}
\begin{array}{c}
sgn_{a} (x-a) \sqrt{\widetilde{g}_a(x-a)} e^{O_a(x-a)+\frac{i}{2 \hbar} E_a(x-a)}=\\
\\
=sgn_{b} (x-b) \sqrt{\widetilde{g}_b(x-b)} e^{O_b(x-a)+\frac{i}{2 \hbar} E_b(x-a)}~,
\end{array}
\label{eqProfile18}
\end{equation}
for almost all $x \in \mathbb{R}^d$.

\vspace{0.2 cm}
\item We have the identity:
\begin{equation}
\|\widetilde{g}_{\alpha}e^{2O_{\alpha}} \|_{L^1 (\mathbb{R}^d)}=1~,~~\alpha =a,b~.
\label{eqProfile19}
\end{equation}
\end{enumerate}

\end{theorem}

According to Theorem \ref{TheoremProfile1}, the first two conditions ensure that $g_{\alpha}$ are Wigner profiles. Condition 3 is the compatibility condition that guarantees that they are the Wigner profiles of the \textit{same} Wigner function $W \psi$ at two different points. The last condition is a restriction on the odd functions $O_{\alpha}$ that ensure a normalized wave function. The full proof of this theorem technical and will be given elsewhere \cite{Dias7}.

To see the effect of the compatibility condition (\ref{eqProfile18}), consider the following example:
\begin{example}\label{Example1}
Suppose that we have the following profile at $x=a$:
\begin{equation}
g_a(p)=N_a e^{-(p-p_a) \cdot M_a (p-p_a)}~,
\label{eqProfile20}
\end{equation}
where $N_a$ is a real constant, $p_a \in \mathbb{R}^d$ and $M_a$ is a real, symmetric, positive-definite $d \times d$ matrix. In view of (\ref{eqProfile17}), we have:
\begin{equation}
N_a \leq \left(\frac{1}{\pi \hbar} \right)^d~.
\label{eqProfile21}
\end{equation}
From the compatibility condition (\ref{eqProfile18}), we conclude that the profile at $x=b$ must satisfy:
\begin{equation}
\begin{array}{c}
sgn_{a} (x-a) \sqrt{N_a}\left(\frac{\pi^d}{\det M_a}\right)^{1/4} \exp\left[-\frac{1}{2 \hbar^2} (x-a) \cdot M_a^{-1} (x-a)+\right.\\
\\
\left. +O_a(x-a)+\frac{i}{\hbar} \left(p_a \cdot(x-a)+ \frac{1}{2} E_a(x-a)\right) \right]=\\
\\
=sgn_{b} (x-b) \sqrt{\widetilde{g}_b(x-b)} e^{O_b(x-a)+\frac{i}{2 \hbar} E_b(x-a)}~,
\end{array}
\label{eqProfile22}
\end{equation}
for almost all $x \in \mathbb{R}^d$.

Taking the modulus squared in the previous equation, we obtain:
\begin{equation}
\begin{array}{c}
A_b(x-b)=\left| \widetilde{g}_b(x-b) \right| =N_a \sqrt{\frac{\pi^d}{\det M_a}} \times \\
\\
\times \exp\left[-\frac{1}{ \hbar^2} (x+b-a) \cdot M_a^{-1} (x+b-a)+2O_a(x-a)-2O_b(x-b)\right]~,
\end{array}
\label{eqProfile23}
\end{equation}
for almost all $x \in \mathbb{R}^d$.

Thus $\widetilde{g}_b $ must be of the form:
\begin{equation}
\begin{array}{c}
 \widetilde{g}_b(x)  =N_a \sqrt{\frac{\pi^d}{\det M_a}} \exp\left[-\frac{1}{ \hbar^2} (x+b-a) \cdot M_a^{-1} (x+b-a)+\right.\\
 \\
 \left. + 2O_a(x-a)-2O_b(x-b) + \frac{i}{\hbar} B_b(x) \right]~,
\end{array}
\label{eqProfile24}
\end{equation}
where $O_a$, $O_b$ are two odd functions and $B_b$ is an even function.

Two conclusions can be drawn from (\ref{eqProfile24}). The first is that, because of the integrability condition (\ref{eqProfile17}), the dominant term (as $|x| \to \infty$) is the quadratic term. The second conclusion is that $\widetilde{g}_b(x) $ has no zeros. 

Thus, for instance, if we choose for $d=1$, the profile
\begin{equation}
g_0(p)=W \psi(0,p)= \frac{e^{- p^2 / \hbar}}{\pi \hbar}~,
\label{eqProfile25}
\end{equation}
at $x=0$, then this is incompatible with a profile
\begin{equation}
g_{\sqrt{\hbar}} (x)=W \psi(\sqrt{\hbar},p)=\frac{e^{-p^2/\hbar}}{\pi \hbar e}  \left(1+ \frac{2 p^2}{\hbar}\right)~.
\label{eqProfile26}
\end{equation}
Both functions (\ref{eqProfile25}) and (\ref{eqProfile26}) are \textit{bona fide} profiles. In fact $g_0(p)$ is the profile (at $x=0$) of the Wigner function associated with the ground state of the simple harmonic oscillator\footnote{We choose mass and frequency $m= \omega=1$ for simplicity.}:
\begin{equation}
W \psi_0(x,p)= \frac{1}{\pi \hbar} e^{- \frac{x^2}{\hbar}- \frac{p^2}{\hbar}}~, 
\label{eqProfile27}
\end{equation}
where as $g_{\sqrt{\hbar}} (x)$ is the profile at $x=\sqrt{\hbar}$ of the Wigner function associated with the first excite state:
\begin{equation}
W \psi_1(x,p)= \frac{1}{\pi \hbar}\left(\frac{2x^2}{\hbar}+\frac{2p^2}{\hbar}-1 \right) e^{- \frac{x^2}{\hbar}- \frac{p^2}{\hbar}}~.
\label{eqProfile28}
\end{equation}
The incompatibility comes from the fact that
\begin{equation}
\widetilde{g}_{\sqrt{\hbar}} (x)=\frac{2}{e \sqrt{\pi \hbar}}\left(1- \frac{x^2}{\hbar}\right) e^{-x^2 / \hbar}
\label{eqProfile28}
\end{equation}
vanishes at $x=\pm \sqrt{\hbar}$ in contradiction with (\ref{eqProfile24}).
\end{example}

\subsection{Solving the Wigner-Moyal equation in an interval with profiles as boundary conditions}

In the Wigner approach to quantum device modelling one tries to solve the Wigner-Moyal equation (or some variant of it) in the interval $\left[0, L \right]$ and impose conditions which ensure existence and uniqueness of the solution. Typically, one could impose some initial condition,
\begin{equation}
F_{Bulk}(x,p,t=0)=F_0(x,p)~,
\label{eqProfile29}
\end{equation}
and profiles as boundary conditions (inflow boundary conditions) \cite{Dimov1,Frensley}:
\begin{equation}
F_{Bulk}(x=0,p,t)=g_0(p,t)~, ~F_{Bulk}(x=L,p,t)=g_L(p,t)~.
\label{eqProfile30}
\end{equation}
The purpose of the simple example given below is to illustrate that a lot of care is needed to ensure the well-posedness of this problem.

Suppose that the system is a harmonic oscillator,
\begin{equation}
H(x,p)=\frac{p^2}{2}+\frac{x^2}{2}~,
\label{eqProfile31}
\end{equation}
with associated Wigner-Moyal equation
\begin{equation}
\frac{\partial }{\partial t} F(x,p,t)= - p\frac{\partial }{\partial x} F(x,p,t)+x\frac{\partial }{\partial p} F(x,p,t)~. 
\label{eqProfile32}
\end{equation}
Moreover, we shall choose to solve this equation only in the interval  $\left[0, L \right]$, and assume the initial condition and the boundary conditions:
\begin{equation}
F(x,p,t=0)=F_0(x,p)~, ~F(x=0,p,t)=g_0(p,t)~, ~F(x=L,p,t)=g_L(p,t)~.
\label{eqProfile33}
\end{equation}
We will now show that all of this is extremely redundant. We only need to specify one of the three functions $F_0$, $g_0$ and $g_L$. 

It is well known that the solution of (\ref{eqProfile32}) is:
\begin{equation}
F(x,p,t)=F_0 \left(x \cos(t)-p \sin (t), x \sin(t) + p \cos(t) \right)~.
\label{eqProfile34}
\end{equation}
So the problem is completely settled if we provide the initial distribution $F_0$. The resulting function $F(x,p,t)$ will be Wigner function for all times if and only $F_0(x,p)$ is a Wigner function. Alternatively, if we chose to provide the profile $g_0(p,t)$ as our boundary condition, then we would have:
\begin{equation}
g_0(p,t)=F(x=0,p,t)=F_0 \left(-p \sin (t),  p \cos(t) \right)~.
\label{eqProfile35}
\end{equation}
Thus, if we are given the initial distribution, we can obtain the profile $g_0$ via (\ref{eqProfile35}). Conversely, if we are given the profile $g_0$, then the initial distribution is:
\begin{equation}
F_0(x,p)=g_0\left(\sqrt{x^2+p^2},\text{arg} \left(-\frac{x}{p}\right) \right)~.
\label{eqProfile36}
\end{equation}
But this also shows that, if the supplementary condition to the Wigner-Moyal equation (\ref{eqProfile32}) is the profile $g_0$ rather than the initial condition $F_0$, then the function $g_0\left(\sqrt{x^2+p^2},\text{arg} \left(-\frac{x}{p}\right) \right)$ must itself be a Wigner function. This is a much more stringent constraint on the profile than the conditions stated in Theorem \ref{TheoremProfile1}. In particular one would have to check the positivity condition (\ref{eqWignerFunctions5B}) or, alternatively, the pure state condition (\ref{eqWignerFunctions5D}).

Regarding the profile at $x=L$, one would come to analogous conclusions:
\begin{equation}
g_L(p,t)=F(x=L,p,t)=F_0 \left(L \cos(t)-p \sin (t), L \sin(t) + p \cos(t) \right)~.
\label{eqProfile37}
\end{equation}

\section{Approximating the approximation with a Wigner function}\label{Approximation}

Despite all the flaws in Frensley's Wigner function approach to quantum device modeling, it is rather surprising that it seems to work reasonably well. The voluminous amount of work based on his approach is a clear proof of this fact. Given that the exact approach, with a complete Wigner-Moyal equation (including boundary potentials) and a thorough and carefull choice of profiles as inflow boundary conditions, may be extremely difficult to solve, one may be tempted to adopt a pragmatic approach and accept Frensley's solution, call it $F(x,p,t)$. To avert obtaining non-physical results (such as negative probability distributions), one could then seek the Wigner function \textit{closest} to $F(x,p,t)$. 

In \cite{Benjamin} the authors obtained an iterative method for obtaining the Wigner function closest to a given square integrable function on the phase space.

Let us be a bit more specific. Let $(\psi_n)$, $(n=1,2,\cdots)$, be a complete orthonormal basis of $L^2(\mathbb{R}^d)$. Then the non-diagonal Wigner functions $(2 \pi \hbar)^{d/2} W(\psi_n,\psi_m)$ constitute a complete orthonormal basis of $L^2(\mathbb{R}^{2d})$. Thus, if $F \in L^2(\mathbb{R}^{2d})$, then we can write:
\begin{equation}
F(x,p)=\sum_{n,m=1}^{\infty} f_{n,m}  W(\psi_n,\psi_m)~,
\label{eqApproximation1}
\end{equation}
with 
\begin{equation}
\|F\|_{L^2(\mathbb{R}^{2d})}^2 =\frac{1}{(2 \pi \hbar)^d} \sum_{n,m=1}^{\infty} |f_{n,m} |^2~.
\label{eqApproximation2}
\end{equation}
Since $F \in L^2(\mathbb{R}^{2d})$, it can be regarded as the Weyl symbol of a Hilbert-Schmidt operator $\widehat{F}$. If $F$ is real, then $\widehat{F}$ is self-adjoint and it has a real countable spectrum. The positive eigenvalues can be written as a decreasing sequence
\begin{equation}
\lambda_1 \geq \lambda_2 \geq \cdots >0~,
\label{eqApproximation3}
\end{equation} 
while the negative eigenvalues can be written as an increasing sequence
\begin{equation}
\lambda_{-1} \leq \lambda_{-2} \leq \cdots <0~.
\label{eqApproximation4}
\end{equation} 
By the spectral theorem, one obtains upon diagonalization in the Weyl-Wigner representation:
\begin{equation}
F(x,p)= \sum_{j=1}^{\infty} \lambda_j W \phi_j + \sum_{j=1}^{\infty} \lambda_{-j} W \phi_{-j}~.
\label{eqApproximation5}
\end{equation} 
Here $\phi_j$, $j \in \mathbb{Z} \backslash \left\{0\right\}$, is a set of orthonormal eigenvectors of $\widehat{F}$ with eigenvalues $\lambda_j$.

The first sum is called the \textit{positive part}
\begin{equation}
F_+(x,p)= \sum_{j=1}^{\infty} \lambda_j W \phi_j ~,
\label{eqApproximation6}
\end{equation} 
and the second sum is called the \textit{negative part}
\begin{equation}
F_-(x,p)= \sum_{j=1}^{\infty} \lambda_{-j} W \phi_{-j} ~.
\label{eqApproximation7}
\end{equation} 
In \cite{Benjamin} we addreessed the following problem. Given a real function $F(x,p) \in L^2(\mathbb{R}^{2d})$, what is the Wigner function $W \psi (x,p) $ closest to $F$ with respect to the $L^2 $ norm? Thus, we wanted to obtain $W \psi$ such that:
\begin{equation}
\|F-W \psi\|_{L^2(\mathbb{R}^{2d})}= \text{min} \|F-W \phi\|_{L^2(\mathbb{R}^{2d})}~,
\label{eqApproximation8}
\end{equation} 
where the minimum is taken over all $\phi \in L^2(\mathbb{R}^{d})$ with $\|\phi \|_{L^2(\mathbb{R}^{d})}=1$.

The result is:
\begin{equation}
W \psi (x,p)= W \phi_1 (x,p)~,
\label{eqApproximation9}
\end{equation} 
where $\phi_1$ is the eigenvector associated with the largest eigenvalue $\lambda_1$ of $\widehat{F}$. Notice that this may not be unique, if the spectrum is degenerate. In that case any linear combination of eigenvectors in the eigenspace of $\lambda_1$ will serve our purposes.

Alternatively, one may be interested in approximating $F$ with a mixed state $W \rho (x,p)$ rather than a pure state $W \psi(x,p)$. In that case, the optimal solution is the positive part of $F$:
\begin{equation}
W \rho(x,p)=F_+(x,p)=\sum_{j=1}^{\infty} \lambda_j W \phi_j ~.
\label{eqApproximation10}
\end{equation} 
This is all very nice, but in general, we may be faced with a difficult task. If the spectrum of $\widehat{F}$ is infinite, then we would have to determine the entire infinite spectrum to find out which is the largest eigenvalue or which is the positive part. 

In \cite{Benjamin} we developed an iterative method, which allows us to obtain the optimal solution with any degree of accuracy. 

Choose an arbitrary (small) $\epsilon >0$ and let $N =N(\epsilon) \in \mathbb{N}$ be  such that
\begin{equation}
\|F-F^{(N)}\|_{L^2 (\mathbb{R}^{2d})} < \epsilon~,
\label{eqApproximation11}
\end{equation} 
where
\begin{equation}
F^{(N)}(x,p)= \sum_{n,m=1}^N f_{n,m} W (\psi_n, \psi_m) (x,p)~.
\label{eqApproximation12}
\end{equation} 
Next compute the spectrum of the finite $N \times N$ matrix $\mathbb{F}^{(N)}= \left[f_{n,m}\right]$, $n,m=1, \cdots, N$. Let $\lambda_1^{(N)} \geq \lambda_2^{(N)} \geq \cdots>0$ be the positive eigenvalues and $\lambda_{-1}^{(N)} \leq \lambda_{-2}^{(N)} \leq \cdots<0$ the negative eigenvalues. It can be shown that $\lambda_j^{(N)}$ increases monotonically to $\lambda_j$, as $N \to \infty$. and $\lambda_{-j}^{(N)}$ decreases monotonically to $\lambda_{-j}$, as $N \to \infty$. More specifically:
\begin{equation}
|\lambda_j^{(N)}-\lambda_j|< 2 (2 \pi \hbar)^{d/2} \epsilon~.
\label{eqApproximation13}
\end{equation}
If we compute the eigenvector $\vec{c}_1^{~(N)}=\left(c_{11}^{(N)},c_{12}^{(N)}, \cdots, c_{1N}^{(N)}\right)$ of the matrix $\mathbb{F}^{(N)}$ associated with its largest eigenvalue $\lambda_1^{(N)}$, then we can compute the corresponding Wigner function:
\begin{equation}
W \psi_1^{(N)} (x,p)= \sum_{n,m=1}^N c_{1n}^{(N)} c_{1m}^{(N) \ast} W (\psi_n, \psi_m) (x,p)~.
\label{eqApproximation14}
\end{equation}
If $W \psi(x,p)$ is the optimal solution, then one can show that:
\begin{equation}
\|W \psi_1^{(N)}-W\psi\|_{L^2(\mathbb{R}^{2d})}\leq 4 \epsilon + \mathcal{O}\left(\epsilon^{3/2} \right)~.
\label{eqApproximation15}
\end{equation}
If we want to approximate $F$ with a mixed state $W \rho$, then one proceeds accordingly to obtain iteratively the positive part $F_+$.

\section{Conclusions}

In this paper we considered the problem of modelling a quantum device using a local formulation based on the Wigner function. More precisely we studied the $*$-genvalue equation and the Moyal dynamical equation for Wigner functions restricted to an interval and satisfying some suitable boundary conditions. 

Since the Wigner function is an intrinsically non-local object, we expected this formulation to display some subtle problems. 

We discussed two different approaches to this problem; one is defined in terms of the Wigner function of the projected wave function into the bulk, the other uses instead the projection of the total Wigner function. While these two objects yield the same probability distribution for the position of the quantum particle in the bulk, they may yield different physical predictions for a generic observable.

We discussed several features of these two formulations.


\begin{enumerate}

\item We showed that the relation between the boundary conditions satisfied by the projected wave function and by the corresponding Wigner function is not straightforward. In particular, the Wigner functions will always satisfy Dirichlet boundary conditions, and the boundary conditions imposed on the wavefunction manifest themselves only at the level of the derivatives of the Wigner function. 

\item Using the simple model of a particle confined to an interval, we also showed that the Wigner function of a projected wave function that satisfys a given time-independent Schr\"odinger equation, does not satisfy the stargenvalue equation that it was supposed to. 
The correct stargenvalue equation has to include new distributional boundary potentials. These boundary potentials are also necessary in the dynamical Wigner-Moyal equation.

\item In the second approach, on the other hand, one uses the restriction of the Wigner function to the bulk, $F_{Bulk}(x,p)$. However, this object is {\bf never} a Wigner function, since it violates the positivity condition. As a result one may obtain negative probability distributions for specific observables.

 
\item The transparent boundary conditions should mimic the effect of the outside Wigner function. They are implemented as appropriate profiles at $x=0$ and $x=L$, commonly known as \textit{inflow boundary conditions}. There is plenty of freedom to choose a profile at one point, say $x=0$. However, if we want to impose a profile at another point $x=L$, then the two profiles have to satisfy a compatibility condition. Also, if we try to impose certain inflow boundary conditions and a certain initial condition for the dynamical Wigner-Moyal equation, then the various conditions may be mutually conflicting leading to an ill-posed problem.

\item The object $F_{Bulk}(x,p)$ is not a Wigner function. However, one may still obtain a true Wigner function that is closest to $F_{Bulk}(x,p)$ with respect to the $L^2$ norm, as explained in section \ref{Approximation}.

\end{enumerate}

An interesting topic that we have only slightly addressed here is how to translate a general boundary condition on the wave function into the Wigner formulation. This is far from being well understood, and it would be a significative development. A correspondence formula would allow us to identify a large set of physical profiles, including those that correspond to self-adjoint boundary conditions. These boundary conditions are important because they identify the domains for which the Hamiltonian operator is self-adjoint. Note that only (essentially) self-adjoint Hamiltonians exhibit a unique unitary time evolution thus preserving probabilities. Of course, by imposing self-adjoint boundary conditions we will further restrict the type of profiles that can be considered for Wigner functions. 

On the other hand, the use of non self-adjoint Hamiltonians has been considered in models of dissipative/dechoerence processes where probabilities are not necessarily preserved \cite{Bender,Weiss}. Such Hamiltonian operators may display complex potentials and/or be associated to non self-adjoint boundary conditions, and are certainly interesting to consider in the context of quantum devices. Once again, a correspondence formula from wave function boundary conditions into the Wigner formulation would allow us to directly generate suitable non self-adjoit profiles.       

Before we conclude, let us briefly mention that, since the non-diagonal term is the cause of $F_{Bulk}$ not being a Wigner function, then one could consider using quasi-distributions other than the Wigner function, which have smaller (negligible) interference terms. This is reported to happen for the Choi-Williams distribution \cite{Cohen1,Cohen2} or the Born-Jordan distribution \cite{Cohen1,Gosson}.

**************************************************************

Author's addresses:

Nuno Costa Dias and Jo\~{a}o Nuno Prata: Grupo de F\'{\i}sica
Matem\'{a}tica, Departamento de Matem\'{a}tica, Faculdade de Ci\^{e}ncias,
Universidade de Lisboa, Campo Grande, Edif\'{\i}cio C6, 1749-016 Lisboa,
Portugal and Escola Superior N\'{a}utica Infante D. Henrique. Av. Eng.
Bonneville Franco, 2770-058 Pa\c{c}o d'Arcos, Portugal

\end{document}